\documentclass[pdflatex,sn-mathphys-num]{sn-jnl}% Math and Physical Sciences Numbered Reference Style 
\usepackage{graphicx}%
\usepackage{multirow}%
\usepackage{amsmath,amssymb,amsfonts}%
\usepackage{amsthm}%
\usepackage{mathrsfs}%
\usepackage[title]{appendix}%
\usepackage{xcolor}%
\usepackage{textcomp}%
\usepackage{manyfoot}%
\usepackage{booktabs}%
\usepackage{algorithm}%
\usepackage{algorithmicx}%
\usepackage{algpseudocode}%
\usepackage{listings}%
\usepackage[export]{adjustbox}
\begin{document}

\title[Large-scale simulations of vortex Majorana zero modes in topological crystalline insulators]{Large-scale simulations of vortex Majorana zero modes in topological crystalline insulators}

\author[1]{\fnm{Chun Yu} \sur{Wan}}

\author[1]{\fnm{Yujun} \sur{Zhao}}

\author[2,3,4]{\fnm{Yaoyi} \sur{Li}}

\author[2,3,4,5,6]{\fnm{Jinfeng} \sur{Jia}}

\author*[1]{\fnm{Junwei} \sur{Liu}}\email{liuj@ust.hk}

\affil[1]{\orgdiv{Department of Physics}, \orgname{Hong Kong University of Science and Technology}, \orgaddress{\street{Clear Water Bay}, \city{Hong Kong}, \country{China}}}

\affil[2]{\orgdiv{Key Laboratory of Artificial Structures and Quantum Control (Ministry of Education), Tsung-Dao Lee Institute, School of Physics and Astronomy}, \orgname{Shanghai Jiao Tong University}, \orgaddress{\city{Shanghai}, \postcode{200240}, \country{China}}}

\affil[3]{\orgname{Hefei National Laboratory}, \orgaddress{\city{Hefei}, \postcode{230088}, \country{China}}}

\affil[4]{\orgname{Shanghai Research Center for Quantum Sciences}, \orgaddress{\city{Shanghai}, \postcode{201315}, \country{China}}}

\affil[5]{\orgname{Southern University of Science and Technology}, \orgaddress{\city{Shenzhen}, \postcode{518055}, \country{China}}}

\affil[6]{\orgname{Quantum Science Center of Guangdong-Hong Kong-Macao Greater Bay Area (Guangdong)}, \orgaddress{\city{Shenzhen}, \postcode{518045}, \country{China}}}

\abstract{Topological crystalline insulators are known to support multiple Majorana zero modes (MZMs) at a single vortex, their hybridization is forbidden by a magnetic mirror symmetry $M_T$. Due to the limited energy resolution of scanning tunneling microscopes and the very small energy spacing of trivial bound states, it remains challenging to directly probe and demonstrate the existence of multiple MZMs. In this work, we propose to demonstrate the existence of MZMs by studying the hybridization of multiple MZMs in a symmetry breaking field. The different responses of trivial bound states and MZMs can be inferred from their spatial distribution in the vortex. However, the theoretical simulations are very demanding since it requires an extremely large system in real space. By utilizing the kernel polynomial method, we can efficiently simulate large lattices with over $10^8$ orbitals to compute the local density of states which bridges the gap between theoretical studies based on minimal models and experimental measurements. We show that the spatial distribution of MZMs and trivial vortex bound states indeed differs drastically in tilted magnetic fields. The zero-bias peak elongates when the magnetic field preserves $M_T$, while it splits when $M_T$ is broken, giving rise to an anisotropic magnetic response. Since the bulk of SnTe are metallic, we also study the robustness of MZMs against the bulk states, and clarify when can the MZMs produce a pronounced anisotropic magnetic response.}

\keywords{Majorana zero mode, Topological crystalline insulator, Topological phase transition, Vortex bound state}

\maketitle

\section{Introduction}\label{sec1}

Coupling MZMs in a controllable way is necessary for realizing fusion or braiding operations in topological quantum computation \cite{kitaev2003fault,alicea2012new,sarma2015majorana,aasen2016milestones,lutchyn2018majorana,beenakker2020search,awoga2024controlling}.
Most topological superconductors realized in experiments \cite{mourik2012signatures, das2012zero, nadj2014observation, xu2015experimental, yin2015observation, ruby2015end, menard2017two, zhang2018observation, wang2018evidence, liu2018robust, machida2019zero, manna2020signature, kezilebieke2020topological,fan2021observation} only feature one MZM at each end, the MZMs from different ends can be coupled via junctions and/or gating \cite{fu2008superconducting,alicea2011non,pikulin2021protocol}.
However, some proposed designs are difficult to implement since the electric field may be screened by the superconducting substrate \cite{zhou2022fusion}.
Alternatively, it is theorized that multiple MZMs can be realized in a single vortex in topological crystalline insulators \cite{CF14,liu2014demonstrating}, the hybridization between MZMs at each end is prohibited by a magnetic mirror symmetry $M_T$. The coupling between MZMs can be controlled by an external field that breaks $M_T$ \cite{CF14}. In particular, the topological crystalline insulator SnTe \cite{hsieh2012topological,tanaka2012experimental,liu2013two,wang2013nontrivial,wang2016electronic,wang2018evidence} is a promising platform since its superconductivity has been realized by proximity effect \cite{klett2018proximity,trimble2021josephson,rachmilowitz2019proximity,yang2019superconductivity,yang2020multiple,liu2024fermi} and doping \cite{hulm1968superconducting, hein1969critical, erickson2009enhanced, balakrishnan2013superconducting, zhong2013optimizing, sato2013fermiology, maurya2014superconducting, maeda2017spin, smylie2018superconductivity, bliesener2019superconductivity, nomoto2020microscopic, smylie2020nodeless, smylie2022full}. Although this magnetic mirror symmetry protection is first proposed in semiconductor nanowires with proximity-induced superconductivity \cite{tewari2012topological}, the topological phase with two MZMs is only stable at some fine-tuned range of parameters \cite{hell2017two}, and has not been identified experimentally. 

However, the existence of crystal-symmetry-protected MZMs cannot be directly probed in experiments, the MZMs are obscured by low-lying vortex bound states \cite{sun2017detection}. And vortices in pristine SnTe could be close to a vortex phase transition \cite{hosur2011majorana,chiu2011vortex} as the bulk states are not gapped \cite{tanaka2012experimental}, MZMs from opposite surfaces may diffuse into the bulk and hybridize. The energy spacing of vortex bound states is approximately $\Delta_0^2/E_F$ \cite{caroli1964bound,volovik1999fermion,khaymovich2009vortex}, where $\Delta_0$ is the superconducting gap at zero field and $E_F$ is the Fermi energy. This energy spacing is about $10\ \text{\textmu}$eV for conventional superconductors \cite{sun2017detection}, which cannot be resolved in experiments, the observed zero-bias peaks (ZBPs) are usually due to non-zero energy vortex bound states \cite{hess1990vortex}.
Nevertheless, in the scanning tunneling microscopy and spectroscopy (STM/STS) study of topological insulator/superconductor heterostructure $\text{Bi}_2\text{Te}_3/\text{Nb}\text{Se}_2$, Xu et al. observed the local density of states (LDOS) changes from a V shape to a Y shape as the chemical potential decreases for samples with increasing thickness \cite{xu2015experimental}, consistent with the emergence of an MZM in the vortex phase transition \cite{li2014majorana,kawakami2015evolution}. However, most previous theories of vortex MZMs omitted the bulk states \cite{fu2008superconducting,liu2014demonstrating} or uses a small lattice that cannot simulate the spatial dependence of the MZMs and the trivial vortex bound states \cite{hosur2011majorana,chiu2011vortex,CF14,yan2020vortex,kobayashi2020double,kobayashi2023crystal}.

We also study how the crystal-symmetry-protected MZMs in the SnTe material class are affected by the bulk states in vortex phase transitions. Previous theories conclude the existence of two MZMs in a vortex in SnTe mainly by analysing the topological surface states with small external fields \cite{CF14,liu2014demonstrating}. Although MZMs can still exist in the presence of metallic bulk states up to a critical chemical potential $\mu=\mu_c$ at which a vortex phase transition occurs \cite{hosur2011majorana,chiu2011vortex,CF14,yan2020vortex,kobayashi2020double,hu2022competing,kobayashi2023crystal}, when the MZMs from opposite surfaces penetrate the bulk and annihilate, the number of MZMs may change as $2\rightarrow1\rightarrow0$ if there is no symmetry that guarantees the degeneracy of vortex line states \cite{kobayashi2020double}.
This is because the symmetry $M_T$ effectively promotes the vortex in SnTe to the Altland-Zirnbauer class BDI with the number of MZMs classified by $\mathbb{Z}$ \cite{tewari2012topological,CF14,kobayashi2020double}, in contrast to $\mathbb{Z}_2$ for 1D topological superconductors in class D \cite{altland1997nonstandard,schnyder2008classification,chiu2016classification}.

In this paper, we study response of the crystal-symmetry-protected MZMs in SnTe class materials in tilted magnetic fields, we simulate the LDOS and study the competition between MZMs and bulk states using kernel polynomial methods (KPM) \cite{weisse2006kernel,nagai2012efficient}. This method allows the simulation of Hamiltonian with dimension larger than $10^8$, bridging the gap between theoretical studies based on minimal models and experimental measurements. For a vortex in superconducting SnTe, we find that the ZBP strongly depends on the direction of the tilted magnetic field. For Fermi level close to the Dirac points, the ZBP elongates when the in-plane component of the field $B_{/\hspace{-1mm}/}$ is along [110] preserving $M_T$, but it splits when $B_{/\hspace{-1mm}/}$ is along [100] breaking $M_T$. At low chemical potential when all MZMs are hybridized, the ZBP splits for both directions. As the spatial distribution of the MZMs differs drastically under tilted magnetic fields \cite{xiong2017anisotropic,kobayashi2019majorana,yamazaki2021magnetic,kobayashi2021majorana,kobayashi2024electromagnetic,yamazaki2024majorana}, the anisotropic magnetic response can demonstrate the existence of symmetry-protected MZMs in experiments \cite{liu2024signatures}.

And we study the vortex phase transitions and evaluate the number of MZMs in tilted magnetic fields. We confirm that in an out-of-plane magnetic field, the number of MZMs changes as $0\rightarrow 2$ as the chemical potential increases, in agreement with ref. \citealp{CF14}. And we point out the single MZM phase is prohibited by a magnetic glide symmetry $G_T$ and the particle-hole symmetry $C$. We then study the topological phase diagram in a tilted magnetic field which may break $M_T$ and/or $G_T$ and changes the classification of the vortex. 

\section{Kernel polynomial method}\label{sec2}
Previous studies simulate the LDOS of vortex states by expanding the radial part of the Bogoliubov quasiparticle amplitudes in Bessel series \cite{shore1989density,gygi1991self,kawakami2015evolution}. However, the efficiency relies on the cylindrical symmetry of the vortex states. Since a tilted magnetic field breaks the cylindrical symmetry and couples bases with different angular momenta, limiting the applicability of this numerical method. In this study we use KPM \cite{weisse2006kernel,nagai2012efficient} to compute the LDOS. Instead of calculating LDOS from the eigenstates and eigenenergies, we directly compute the LDOS in KPM by expanding the Green's functions in Chebyshev moments. 

The retarded Green's function is given by \cite{datta1997electronic}
\begin{equation}\label{eq_b1}
	G^R(E)=\frac{1}{E-H+\text{i}\eta},
\end{equation}
which can also be represented by the eigenenergies $E_k$ and the eigenstates $| k \rangle$,
\begin{equation}\label{eq_b2}
	G^R(E)=\sum_k\frac{| k \rangle \langle k | }{E-E_k+\text{i}\eta}.
\end{equation}
The LDOS for the $i^{th}$ basis is defined as
\begin{equation}\label{eq_b_dos}
	d^i(E)=\sum_k\langle i | k \rangle \langle k | i \rangle \delta(E-E_k),
\end{equation}
in the following sections the LDOS is summed over different spins and orbitals, we omit the summation below to simplify the notations. Comparing with eq. \ref{eq_b2} we have
\begin{equation}\label{eq_b4}
	d^i(E)=-\frac{1}{\pi}\langle i |\text{Im}(G^R)| i \rangle.
\end{equation}
where the Dirac delta function is replaced by a Lorentz function. And the imaginary part of the energy $\eta$ determines the spectral width of the LDOS.

In KPM, the Green's functions are expanded in terms of Chebyshev moments, in this paper we focus on the diagonal parts which are the LDOS. First, since Chebyshev polynomials are defined in the range $[-1,1]$, the Hamiltonian and the energy are rescaled by the maximum energy $E_{\text{max}}$, $\tilde{H}=H/E_{\text{max}}$ and $\tilde{E}=E/E_{\text{max}}$.
We want to expand the LDOS in terms of Chebyshev moments, which are defined as
\begin{equation}\label{kpm3}
	\mu^{i}_n =\int_{-1}^{1}d^i(\tilde{E})T_n(\tilde{E})d\tilde{E},
\end{equation}
where $T_n(\tilde{E})$ is the $n^{th}$ Chebyshev polynomial. Using eq. \ref{eq_b1}, we have
\begin{equation}\label{kpm4}
	\mu^{i}_n=\langle i |T_n(\tilde{H})| i \rangle.
\end{equation}
This moment is calculated recursively using
\begin{equation}\label{kpm5}
	T_{n+1}(\tilde{H})| i \rangle=2\tilde{H}T_n(\tilde{H})| i \rangle-T_{n-1}(\tilde{H})| i \rangle.
\end{equation}
Finally, the LDOS is reconstructed using the moments $\mu_n^i$ and the kernel polynomials $g_n$ which dampens the Gibbs oscillations of the partial sum.
\begin{equation}
	d^{i}(\tilde{E})\approx\frac{1}{\pi\sqrt{1-\tilde{E}^2}}\Bigg[ g_0\mu^i_0+2\sum_{n=1}^{N}g_n\mu^i_n T_n(\tilde{E})\Bigg].
\end{equation}
And we use the Jackson kernels
\begin{equation}
	\hspace{-5mm}g_n=\frac{1}{N+1}\Bigg[ (N-n+1)\text{cos}\big( \frac{\pi n}{N+1} \big) +\text{sin}\big( \frac{\pi n}{N+1} \big)\text{cot}\big( \frac{\pi }{N+1} \big)\Bigg],
\end{equation}
the spectral density has a Gaussian widening with width $\sigma\approx \frac{\pi E_{\text{max}}}{N}$ \cite{weisse2006kernel}. For all KPM calculations in this work, $E_{\text{max}}\approx 5$.

In experiments the resolution is lower than the energy spacing between the vortex bound states, which means a smaller number of moments $N$ and reduces the computation time. Although KPM has been used to study vortex bound states \cite{nagai2012direct,smith2016manifestation,berthod2016vortex,galvis2018tilted,berthod2018signatures}, the systems in these studies are two-dimensional while we need to consider a three-dimensional system to account for the competition between the surface states and the bulk states, the dimension of the Hamiltonian is larger by two orders of magnitude.

\section{Model Hamiltonian}\label{sec3}
\begin{figure}[h]
	\includegraphics[width=0.8\columnwidth]{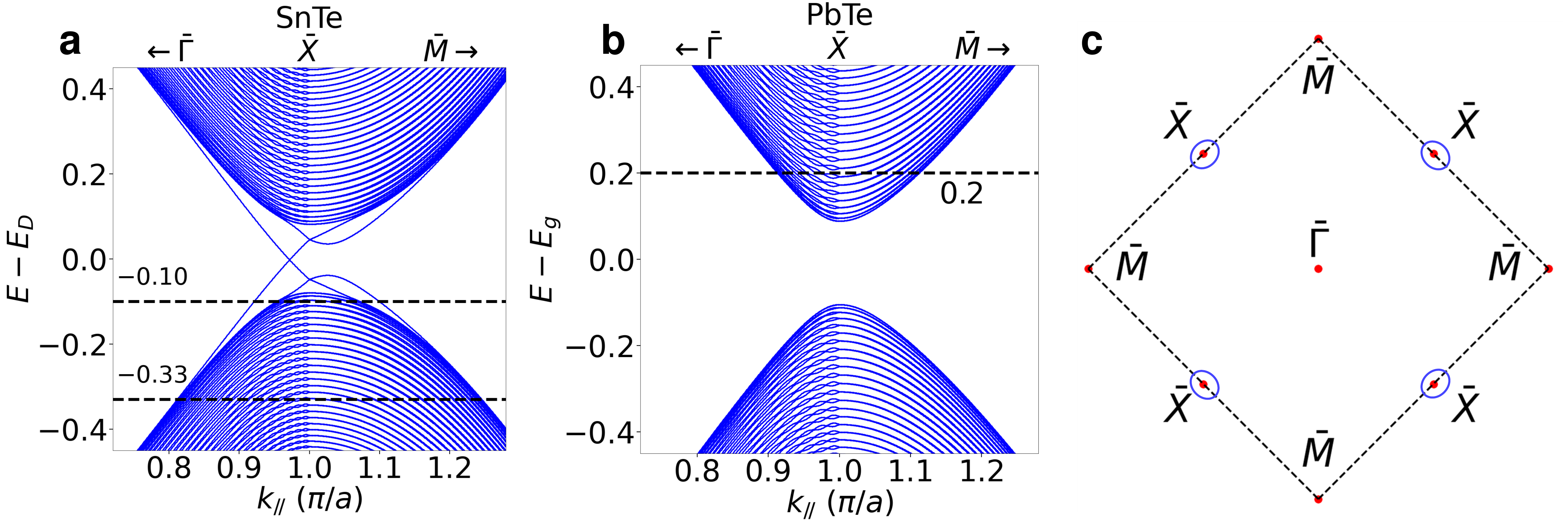}
	\caption{\label{fig1} (001) surface band structures for SnTe (a) and PbTe (b), $E_D=-0.04$ denotes the energy of the Dirac points. In our model for SnTe, the valence band maximum is at $E-E_D=-0.08$. The black dashed lines mark the chemical potential we use to simulate the LDOS. (c) Schematic diagram for the Fermi surface for the (001) surface Brillouin zone.}
\end{figure}

Here we adopt a tight-binding model consisting of 3 $p$-orbitals for Sn (or Pb) and Te atoms respectively \cite{mitchell1966theoretical, hsieh2012topological, fulga2016coupled}.
Our analysis is specific to the $(001)$ surface of SnTe material class which have a rocksalt structure with valence band maximum (or conduction band minimum for PbTe) at $L$ points \cite{hsieh2012topological,liu2013two}.
In real space, the Hamiltonian reads
\begin{equation} \label{band_hamiltonian}
	\begin{split}
		h = &\sum_{j}(m_j-\mu)\sum_{\boldsymbol{r},s}\boldsymbol{c}^\dagger_{js}(\boldsymbol{r})\cdot\boldsymbol{c}_{js}(\boldsymbol{r})  \\
		+ &\sum_{j,j^{\prime}}t_{j,j'}\sum_{(\boldsymbol{r},\boldsymbol{r^\prime}),s}\bigg[\boldsymbol{c}^\dagger_{js}(\boldsymbol{r})\cdot\boldsymbol{d}_{\boldsymbol{r,r^\prime}} \boldsymbol{d}_{\boldsymbol{r,r^\prime}} \cdot \boldsymbol{c}_{j^{\prime}s}(\boldsymbol{r^\prime})+\text{H.c.}\bigg] \\
		+ &\sum_{j}\text{i}\lambda_j\sum_{\boldsymbol{r},s,s^\prime}\boldsymbol{c}^\dagger_{js}(\boldsymbol{r})\times\boldsymbol{c}_{js^\prime}(\boldsymbol{r})\cdot\pmb{\sigma}_{s,s^\prime},
	\end{split}
\end{equation}
where $j=a,\ c$ represents the anion (Te) and the cation (Sn or Pb) respectively. The creation and annihilation operators $\boldsymbol{c}^\dagger_{js}(\boldsymbol{r}), \boldsymbol{c}_{js}(\boldsymbol{r})$ compose of the orbitals $p_x,\ p_y$ and $p_z$. $\boldsymbol{d}_{\boldsymbol{r,r^\prime}}$ represents a unit vector between nearest neighbors $(\boldsymbol{r},\boldsymbol{r^\prime})$. $\lambda_j$ is the atomic spin-orbit coupling strength. $\pmb{\sigma}$ are the Pauli matrices for the spin basis. $\mu$ represents the chemical potential. We choose the parameters $m_c=-m_a=1.65, t_{ac}=t_{ca}=1.2$ and $l_a=l_c=0.5$. The band structure for the $(001)$ surface is plotted in Fig. \ref{fig1}(a), with $t_{cc}=-t_{aa}=0.38$, the system is topologically nontrivial with Dirac cones protected by mirror symmetries $M_{110}$ and $M_{1\bar{1}0}$ as in SnTe. For $t_{cc}=-t_{aa}=0.33$, the system is gapped [Fig. \ref{fig1}(b)], this serves as a model for PbTe which is a trivial insulator. For the SnTe model, the Dirac points are at $E_D=-0.04$, and we consider the chemical potential starting from the Dirac points, denote as $\mu-E_D$, with $\mu-E_D<0$ since SnTe is a p-type semiconductor. The $(001)$ surface Brillouin zone for SnTe is shown in Fig. \ref{fig1}(c), there are four Dirac cones near the $\bar{X}$ points.
We consider superconductivity in SnTe with a single-gap $s$-wave pair potential, and omit the smaller gap related to the topological surface states \cite{bliesener2019superconductivity,yang2020multiple} to simplify the calculations.
We include a superconducting vortex along the $z$ direction,
\begin{equation}
	\Delta(\rho,\theta)=\Delta_0 \text{tanh}(\rho/\xi_0)e^{\text{i}\theta},
\end{equation}
where $\rho=\sqrt{x^2+y^2}$, $\theta=\text{tan}^{-1}(y/x)$, $\Delta_0$ is the superconducting gap and $\xi_0$ is the vortex core size. MZMs (if they exist) appear at the end points on the $(001)$ and $(00\bar{1})$ surfaces.

We consider an additional in-plane magnetic field $B_{/\hspace{-1mm}/}$ in the form of a uniform Zeeman term (eq. \ref{zeeman}) and keep the vortex profile unchanged to retain the translational symmetry in the $z$ direction, this simplifies the calculations for the vortex phase transitions in Sec. \ref{sec5}. In experiments, the Zeeman term is mainly contributed by the orbital effect which has the same form as the spin effect for topological surface states \cite{zhu2021discovery,pan2024majorana}. This approximation neglects the tilting of the vortex line, it is valid for a range below the surface of SnTe that is much shorter than the penetration depth of the superconducting substrate.
\begin{equation}\label{zeeman}
	h_{\text{Zeeman}} = E_Z  \sum_{j,\boldsymbol{r},s,s^\prime}
	\hat{\boldsymbol{B}}_{\parallel}\cdot \pmb{\sigma}_{ss^\prime}
	\boldsymbol{c}^\dagger_{js}(\boldsymbol{r})\cdot\boldsymbol{c}_{js}(\boldsymbol{r}),
\end{equation}
where $E_Z$ is the Zeeman energy and $\pmb{\sigma}$ are the Pauli matrices. In this paper we mainly consider the in-plane component of the magnetic field to be along $[110]$ or $[100]$ direction.

In experiments, identification of vortex MZMs is difficult because the energy levels are too closely packed. In the simulations this can be quantified by $\Delta_0/\delta$, where $\delta$ is the energy of the lowest vortex bound states other than the MZMs. Conventional superconductors have $\Delta_0/\delta\gg1$, then the energy spacing is approximately equal \cite{caroli1964bound,volovik1999fermion,khaymovich2009vortex}, $E\approx n \delta$,
where $n$ is an integer.
And $\delta$ depends on the Fermi energy and the vortex core size \cite{caroli1964bound}.
\begin{equation}
	\delta\approx\frac{\Delta_0}{k_F\xi_0},
\end{equation}
where $k_F$ is the Fermi wavevector.
Since the vortex core size is proportional to the BCS coherence length in most superconductors, $\delta\sim\Delta_0^2/E_F$  \cite{caroli1964bound} as mentioned in the introduction. In the simulations, we treat $\Delta_0$, $\xi_0$ and $E_F$ as independent parameters, and $E_F=\mu-E_D$ for the topological surface states of SnTe. To simulate the LDOS for the vortex bound states, the dimensionless quantity $\Delta_0/\delta \approx k_F\xi_0$ should be large so that the energy spectrum appears continuous as in experiments. Because the LDOS for the vortex bound states should vanish at the boundaries, the lattice must be large, with $N_x\gg\xi_0$ and $N_y\gg\xi_0$.

\section{Local density of states}\label{sec4}
We choose lattices with $300\times300\times50$ atoms and $\xi_0=40$, the Hamiltonian has dimension $D\sim 10^8$ which is difficult to be diagonalized. The time complexity of KPM scales as $O(DN)$ \cite{weisse2006kernel,nagai2012efficient}. In KPM, the Green's functions are expanded as a partial sum involving $N$ Chebyshev moments, and the energy resolution of the LDOS scales as $E_{\text{max}}/N$ for the Jackson kernel \cite{weisse2006kernel}. Since the energy resolution is inversely proportional to the number of iterations ($N/2$), we choose $\Delta_0=0.05$ to reduce computation time. Although $\Delta_0$ is comparable to $E_F$, and makes the LDOS strongly asymmetric about $E_F$, we symmetrize it above and below $E_F$. For tilted magnetic fields, we choose the scanning direction of the LDOS along $B_{/\hspace{-1mm}/}$. For LDOS along either [110] or [100], only the Te atoms are sampled since the focus is on the valence bands.

The quality of the simulations can be checked by examining the ratio $\Delta_0/\delta$ which can be approximated by $k_F\xi_0$. For $\mu-E_D=-0.1$, we have $\Delta_0/\delta=8.22$ [Fig. \ref{fig2}(d)]. Although it is smaller than $k_F\xi_0\sim 100$ for SnTe$/$Pb in the recent experiment \cite{liu2024signatures}, the LDOS has a Y shape [Fig. \ref{fig2}(a)] for $N=2000$ when the individual vortex bound states cannot be resolved, these parameters qualitatively reproduce the LDOS in experiments.

\subsection{Anisotropic magnetic response}\label{sec4a}
\begin{figure}[h]
	\includegraphics[width=0.6\columnwidth]{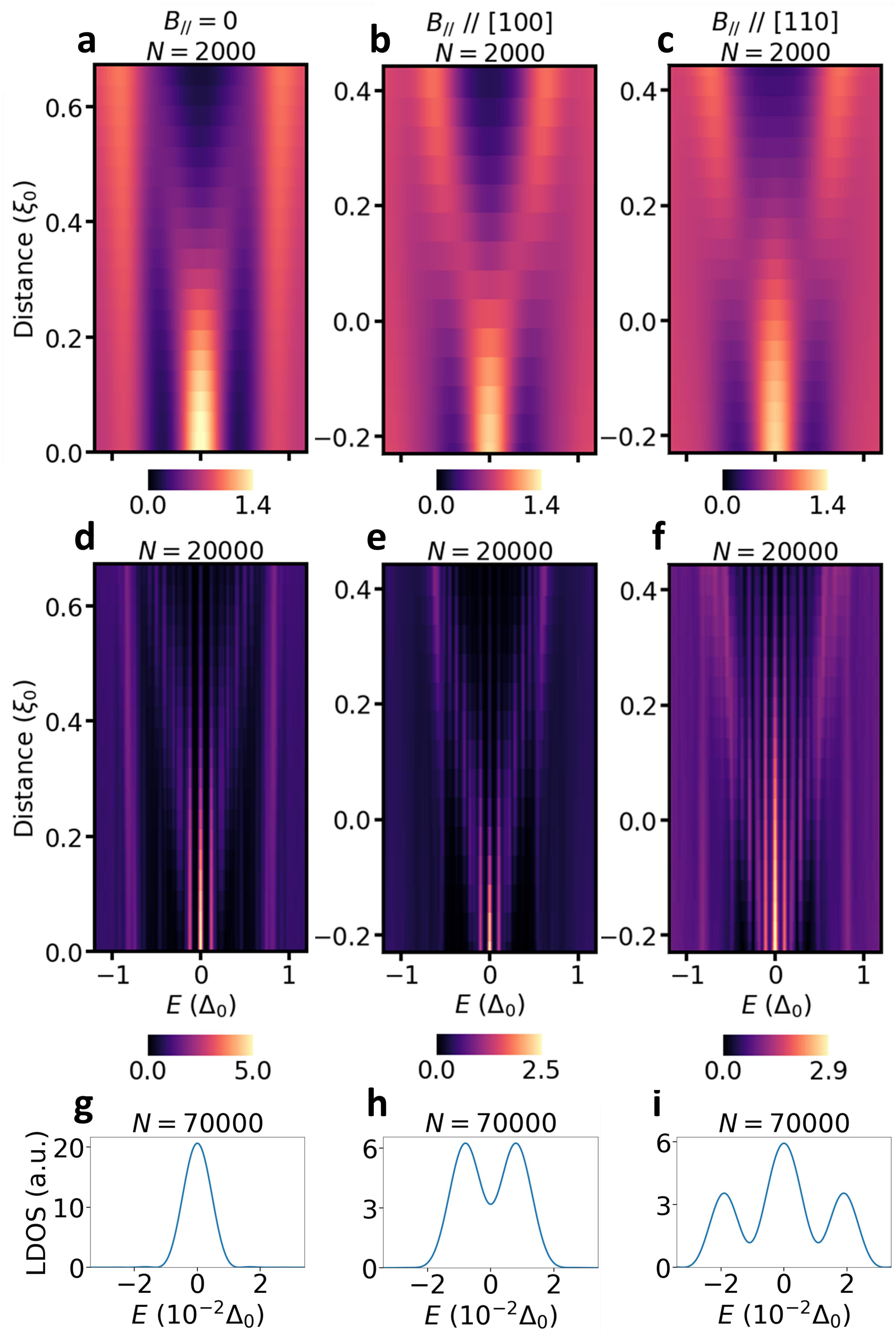}
	\caption{\label{fig2} Vortex LDOS for the (001) surface of SnTe in tilted magnetic fields at different resolutions. MZMs are hybridized in the second column when $B_{/\hspace{-1mm}/}$ is along [100] breaking $M_T$. The resolution is enhanced by increasing the number of Chebyshev moments $N$ in KPM. (a-c) $N=2000$ where the individual vortex bound states are not resolved, the ZBP is sensitive to the direction of $B_{/\hspace{-1mm}/}$. (d-f) LDOS for $N=20000$ and the vortex bound states are resolved. (g-i) LDOS for $N=70000$ at the density peaks, MZMs are absent in (h). The Zeeman energy is $E_Z=0.4\Delta_0$ and $\Delta_0=0.05$.}
\end{figure}

For closely packed energy levels with $\Delta_0/\delta\gg1$, the splitting of the ZBP strongly depends on the direction of $B_{/\hspace{-1mm}/}$ and whether it preserves $M_T$, we call this the anisotropic magnetic response of the vortex bound states. For $E_Z=0.4\Delta_0$, the anisotropic magnetic response is apparent even at $N=2000$. For $B_{/\hspace{-1mm}/}$ along [100] which destroys $M_T$ that protects the MZMs, the ZBP moves towards the opposite direction of $B_{/\hspace{-1mm}/}$, the splitting enlarges and the LDOS resembles a V shape [Fig. \ref{fig2}(b)]. For $B_{/\hspace{-1mm}/}$ along $[110]$, $M_T=TM_{1\bar{1}0}$ is preserved, the ZBP elongates and does not split even away from the vortex core [Fig. \ref{fig2}(c)].

Using $N=20000$, we can resolve the energy spacing between some vortex bound states [Fig. \ref{fig2}(d-f)]. In particular, for $B_{/\hspace{-1mm}/}$ along [110], we can see that the extended ZBP from Fig. \ref{fig2}(f) consists of vortex bound states other than the MZMs. This anisotropic magnetic response originates from the nontrivial spin texture of the topological surface states in SnTe, it is not a direct consequence of destroying or preserving the MZMs.
Further increasing the resolution to $N=70000$ at the sites with the highest LDOS, we confirm that the MZMs are hybridized when $B_{/\hspace{-1mm}/}$ is along [100]. The anisotropic magnetic response is pronounced even at low resolution when the splitting of the MZMs cannot be observed, in Fig. \ref{fig2}(h), the splitting is only $0.90\%$ of the superconducting gap $\Delta_0$ for Zeeman energy $E_Z=0.4\Delta_0$. The hybridization of crystal-symmetry-protected MZMs drastically changes their spatial distribution, our study establishes an experimentally viable approach to probe them in a wide range of topological crystalline materials \cite{po2017symmetry,bradlyn2017topological,kruthoff2017topological,ono2019symmetry,ono2021z}.

\subsection{Comparisons with trivial phases}\label{sec4b}
\begin{figure}[h]
	\includegraphics[width=0.6\columnwidth]{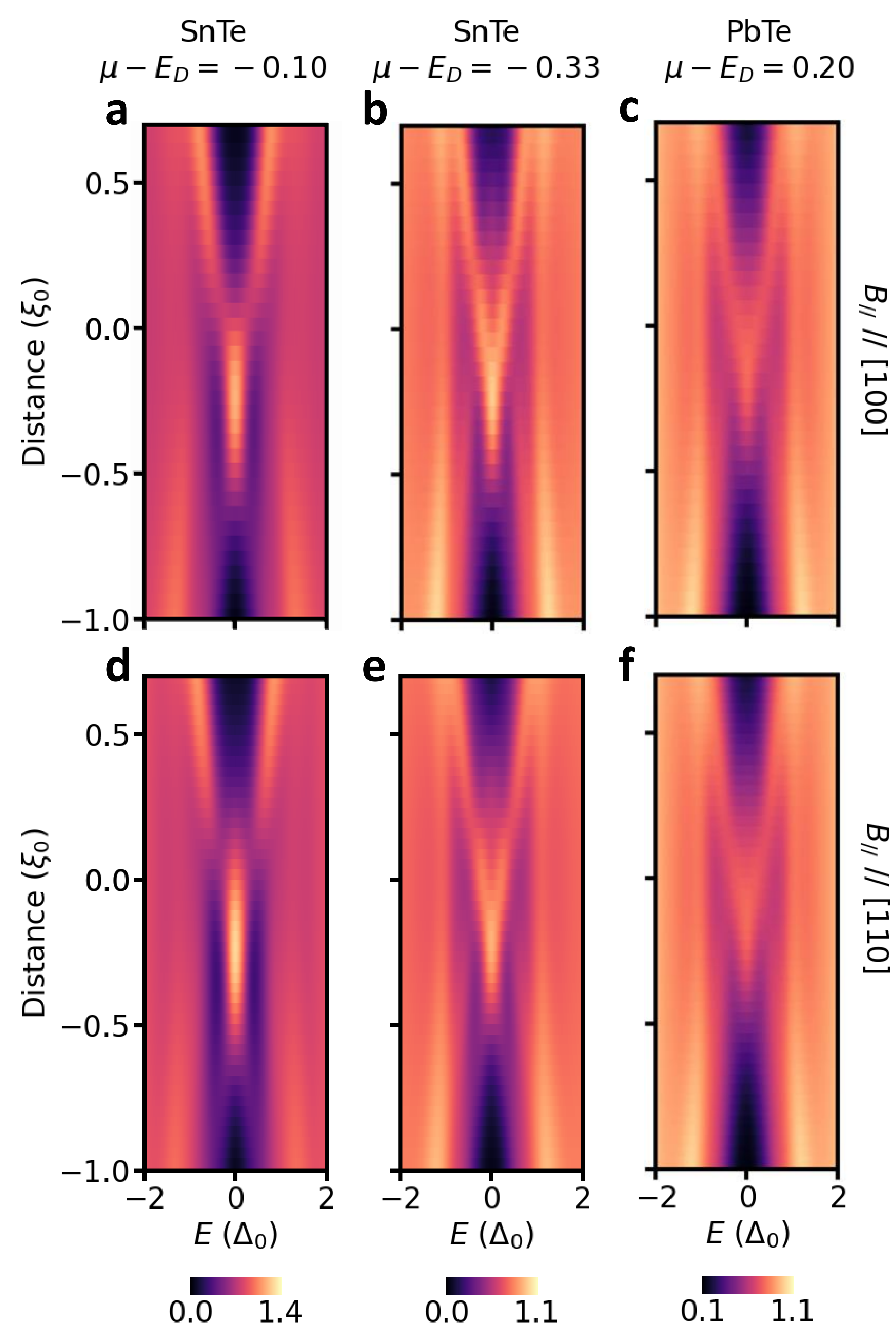}
	\caption{\label{fig3} Vortex LDOS in tilted magnetic field for SnTe and PbTe. (a-c) When $B_{/\hspace{-1mm}/}$ is along [100], any mirror-symmetry-protected MZMs must be hybridized, and the ZBPs split. (d-f) When $B_{/\hspace{-1mm}/}$ is along [110], $M_T$ is preserved and SnTe at low chemical potential supports two MZMs. For SnTe in the topological phase, the ZBP extends much longer than SnTe at low chemical potential as the MZMs diffusive into the bulk and may hybridize. $E_Z=0.4\Delta_0$ for (a, d), and $E_Z=0.3\Delta_0$ for the other subplots.}
\end{figure}

As studied in Sec. \ref{sec5}, the MZMs can be hybridized at low chemical potential $\mu<\mu_c$ in vortex phase transitions without breaking any symmetry, and we find that the magnetic response becomes isotropic at lower $\mu$ when the LDOS is dominated by bulk states. For SnTe at $\mu-E_D=-0.10$, the anisotropic magnetic response [Fig. \ref{fig3}(a,d)] arises because the vortex bound states are mostly contributed by the topological surface states which are sensitive to the direction of $B_{/\hspace{-1mm}/}$. At $\mu-E_D=-0.33$, the MZMs diffuse into the bulk and hybridize (confirmed with KPM in Appendix \ref{appendix2}), the LDOS on the surface is mostly contributed by the bulk bands, the splitting of the ZBP becomes isotropic [Fig. \ref{fig3}(b,e)]. In Appendix \ref{appendix2}, we also show the LDOS at fixed $E_Z$ and various $\mu$.
The change in the magnetic response of the LDOS is gradual and not sensitive to the number of MZMs (Fig. \ref{figF3}).

We further confirm the isotropic magnetic response arises from the bulk states by comparing with PbTe which does not feature topological surface states. The ZBP splits for either directions of $B_{/\hspace{-1mm}/}$ [Fig. \ref{fig3}(c,f)]. Although in Fig. \ref{fig3}(e,f) there are states at distance $>0$ similar to the elongated ZBP in \ref{fig3}(d), their density is weak and the LDOS still has a V shape, the slight anisotropy in the trivial cases could be caused by the underlying cubic lattice. 

\section{Vortex Phase Transitions}\label{sec5}
\subsection{Under out-of-plane magnetic field}\label{sec5a}
We construct the Bogoliubov-de Gennes (BdG) Hamiltonian for the vortex line (eq. \ref{bdg_hamiltonian}) which is periodic in the $z$ direction, and we define the symmetry operators in Appendix B.
\begin{equation}\label{bdg_hamiltonian}
	H_{\text{BdG}}(k_z)=\left(\begin{array}{cc}
		h(k_z) & \Delta(\rho,\theta)i\sigma_y \\
		-{\Delta}^{\ast} (\rho,\theta)i\sigma_y & -h^T(-k_z)
	\end{array}\right),
\end{equation}
where $\Delta(\rho,\theta)=\Delta_0 \text{tanh}(\rho/\xi_0)e^{i\theta}$.

We clarify the number of MZMs in SnTe which features four Dirac cones protected by $M_{110}$ and $M_{1\bar{1}0}$. Since a superconducting vortex breaks mirror symmetries, the system preserves magnetic mirror symmetries $M_T$ instead. Although either magnetic mirror symmetry $M_T=M_{110}T$ or $M_{1\bar{1}0}T$ can protect two MZMs, in an out-of-plane magnetic field when the system preserves both, it still hosts only two MZMs. As the system respects a screw symmetry $S_{4z}$, the Hamiltonian can be block-diagonalized by $S_{4z}$ with eigenvalues $\{1,\text{i},-1,-\text{i}\}$ \cite{qin2019quasi,kobayashi2023crystal}. Since the particle-hole symmetry $C$ exchanges the i and $-$i blocks, only the $1$ and $-1$ blocks can support MZMs which are self-Hermitian. Since $M_T$ commutes with $S_{4z}$, there is chiral symmetry within each block, these two blocks can each support multiple MZMs \cite{kobayashi2020double}. Although this model for SnTe with $S_{4z}$ symmetry is classified by $\mathbb{Z}\times\mathbb{Z}$, we only find two MZMs with our range of parameters. Our conclusions are consistent with refs. \citealp{CF14,liu2014demonstrating} that evaluate the number of MZMs using the Jackiw-Rossi model \cite{jackiw1981zero} for the multiple Dirac cones in SnTe.

\begin{figure}[h]
	\includegraphics[width=0.98\columnwidth]{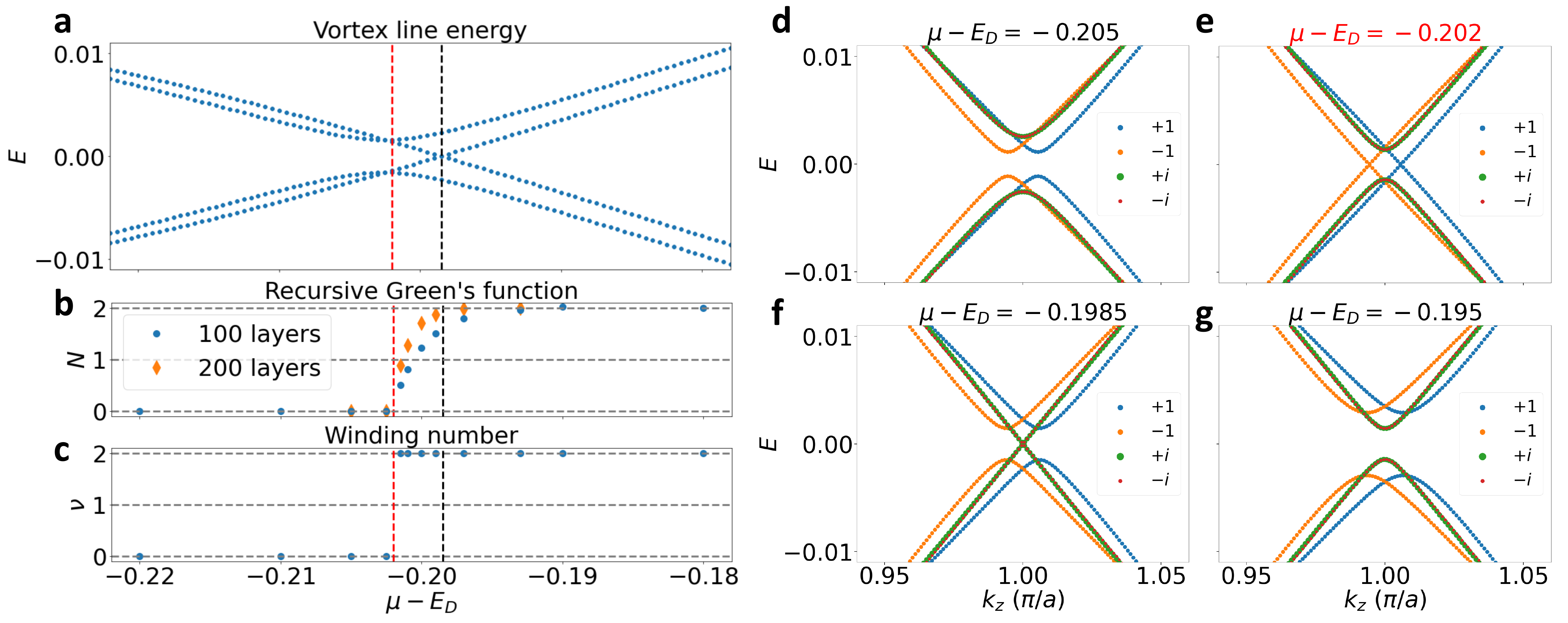}
	\caption{\label{fig4} Vortex phase transition in SnTe in an out-of-plane magnetic field. (a) The vortex line energy at $k_z=\pi/a$, it is at least two-fold degenerate due to $G_T$. The number of MZMs are calculated using recursive Green's functions (b) and winding number (c). The black dots denote crossings at $k_z=\pi/a$, the red dots are for $k_z\neq\pi/a$. (d-g) Vortex line energy for different $k_z$, the colors denote the $S_{4z}$ eigenvalues. We use $\Delta_0=0.1$, $\xi_0=1$ for the vortex and $N_x=N_y=20$ for the lattice.}
\end{figure}

A vortex phase transition is signaled by (1) a crossing of the bulk energy, (2) appearance or disappearance of boundary states and (3) a change in the topological invariant. In Fig. \ref{fig4}(a, d-g), we show the vortex line energy as a function of chemical potential $\mu$. As $\mu$ increases, the Fermi level is closer to the Dirac points and there are two sets of crossings, $\mu_c=-0.242,-0.236$. We use recursive Green's functions \cite{sancho1985highly} to calculate the number of zero energy states and show their distribution near the surface (Appendix C). The system respects a chiral symmetry ($\Gamma=CM_T$, the explicit form is given in Appendix B) and a winding number $\nu$ can be defined (Appendix D), and it corresponds to the number of MZMs at each end of the vortex line \cite{tewari2012topological,kobayashi2020double}. We label the vortex line energy with screw symmetry $S_{4z}$ eigenvalues, the MZMs have eigenvalues $\pm 1$ and they hybridize at the first crossing point [Fig. \ref{fig4}(e)]. The remaining crossings have eigenvalues $\pm$i [Fig. \ref{fig4}(f)] and they do not couple to the MZMs. Since the crossings for eigenvalues $\pm$i occurs at $k_z=\pi/a$ and they overlap, the gap will open again and there are no nodal vortex phases in this model for $\mu-E_D<0$.

Although ref. \citealp{CF14} shows that $M_T$ can protects two MZMs by analyzing the surface Hamiltonian, it is possible that the number of MZMs changes as $2\rightarrow1\rightarrow0$ during the vortex phase transitions \cite{kobayashi2020double}.
Then if we introduce a perturbation that breaks $M_T$, the system changes from the Altland-Zirnbauer class BDI to D which can still host a single MZM \cite{tewari2012topological}. However, this is forbidden if a magnetic glide symmetry $G_T=TG_y$ or $TG_x$ is present. As shown in Fig. \ref{fig4}(a-c), the vortex line energy is at least two-fold degenerate at $k_z=\pi/a$ since $G_T^2=-1$ at that point. And the transition at $k_z$ away from $\pi/a$ must also appear in pairs because of the particle-hole symmetry $C$ which gives $CH(k_z)C^{-1}=-H(-k_z)$. The symmetries $C$, $M_T$ and $G_T$ together forbid an odd number of MZMs. For odd $N_x=N_y$, there is no $G_T$ and a single MZM can be realized (Fig. E1). And $S_{4z}$ is replaced by $C_{4z}$ which does not exchange the sublattices. However, in experiments the vortex size is much larger than the lattice parameter so that there is no difference between even and odd $N_x=N_y$. The $\nu=1$ phase for odd $N_x=N_y$ is an artifact from the small lattice used in the calculations.

\subsection{Under tilted magnetic field}\label{sec5b}
\begin{figure}[h]
	\includegraphics[width=0.98\columnwidth]{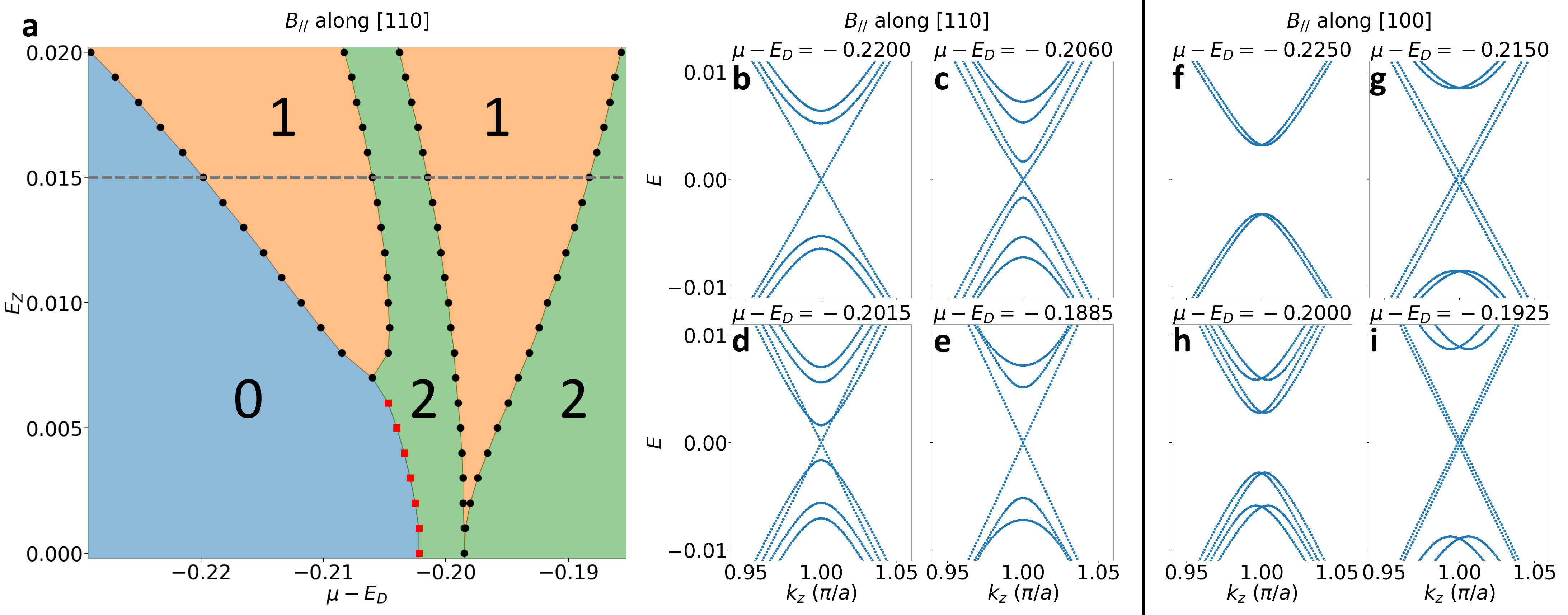}
	\caption{\label{fig5} Vortex phase transition in SnTe in tilted magnetic fields. We include the in-plane component as a Zeeman field and consider it along [110] and [100] directions. The phase diagram of vortex phase transition for $B_{/\hspace{-1mm}/}$ along [110] at various chemical potential $\mu$ and Zeeman energy $E_Z$ (a). The red dots denote a pair of crossings at $(k_z,2\pi/a-k_z)$ for $k_z\neq \pi/a$ enforced by particle-hole symmetry. The four crossings at $E_Z=0.015$ occur at $k_z=\pi/a$ (b-e). For comparison, $TG_y$ is preserved when $B_{/\hspace{-1mm}/}$ is along [100], it ensures the energy at $k_z=\pi/a$ is at least two-fold degenerate (f-i).}
\end{figure}

For $B_{/\hspace{-1mm}/}$ along $[110]$, the system can still host two MZMs as $TM_{1\bar{1}0}$ is preserved. We determine the phase diagram for different $\mu$ and $E_Z$. Since both $TG_x$ and $TG_y$ are broken, the degeneracy at $k_z=\pi/a$ is lifted and the system can host a single MZM in between vortex phase transitions. Without $S_{4z}$, the number of MZMs $\nu$ changes at each crossing point. For $E_Z>0.006$, all crossing points occur at $k_z=\pi/a$ with $\nu$ changes by 1 [Fig. \ref{fig5}(b-e)]. And the regions with $\nu=1$ grows with increasing $E_Z$.

For $B_{/\hspace{-1mm}/}$ along $[100]$, the system cannot host two MZMs as $TM_{110}$ and $TM_{1\bar{1}0}$ are broken. Since $TG_y$ is preserved, the vortex line energy is at least two-fold degenerate at $k_z=\pi/a$ [Fig. \ref{fig5}(f-i)]. $TG_y$ and $C$ ensure $\nu$ can only change by 2 in a vortex phase transition, because $\nu=0$ or $2$ at $E_Z=0$, we have $\nu=0$ for any $\mu$ and $E_Z$.

Even when $M_T$ is preserved, the MZMs can be hybridized if $\mu<\mu_c$.  From the vortex line energy and the winding number we conclude the number of MZMs to be $\nu\in\{0,1,2\}$ for $B_{/\hspace{-1mm}/}$ along $[110]$, and $\nu=0$ for $B_{/\hspace{-1mm}/}$ along $[100]$. The conditions for the $\nu=2$ phase we determined may allow schemes to braid the two MZMs in a single vortex by changing the direction of the magnetic field or other symmetry-breaking fields \cite{pahomi2020braiding,pan2022detecting,liu2024magnetically}.

Although we evaluated the number of MZMs in SnTe using a small lattice model, the LDOS in the last section is calculated using the same lattice model with larger $N_x$, $N_y$ and $\xi_0$, the conclusions for topology of the vortex still applies but the phase transition points $(\mu_c,E_{Zc})$ may vary.

\section{Conclusions}\label{sec6}
We have studied the topological vortex phases of topological crystalline insulator SnTe in tilted magnetic fields. We verified that two MZMs can coexist in a vortex protected by a magnetic mirror symmetry, and we pointed out that a single MZM in between vortex phase transitions is prohibited if a magnetic glide symmetry is preserved. By extending the calculations to larger lattice models, we found that the LDOS displays an anisotropic magnetic response when the vortex features MZMs. It is due to the drastic difference between the spatial distribution of MZMs and trivial vortex bound states. This anisotropic magnetic response remains robust for a range of $\mu$ below the valence band maximum and gradually disappears when MZMs diffuse into the bulk. The elongated ZBP in a tilted magnetic field that preserves $M_T$ is a distinct experimental signature of the crystal-symmetry-protected MZMs in a vortex in a topological crystalline insulator.

\backmatter

\bmhead{Acknowledgements}
The simulations were done on the Hefei advanced computing center and Tianhe-2 in the National Supercomputer Center in Guangzhou.

\bmhead{Funding}
We acknowledge the financial support from the National Key R$\&$D Program of China (Grants No. 2021YFA1401500), the National Science Foundation of China (Grants No. 12022416) and the Hong Kong Research Grants Council (Projects No. N$\_$HKUST626/18, ECS26302118 and 16305019).

\bmhead{Author contribution}
JL supervised the project. CYW and YZ performed the calculations. CYW, YZ, YL, JJ and JL analysed the results. JL and CYW wrote the manuscript with contributions from all authors.
%\end{itemize}

%\noindent
%If any of the sections are not relevant to your manuscript, please include the heading and write `Not applicable' for that section. 


\renewcommand\thefigure{\thesection.\arabic{figure}}    
\begin{appendices}

\section{LDOS at various chemical potential}\label{appendix2}
\setcounter{figure}{0} 
In this section we study how the magnetic response of the LDOS changes as $\mu$ is varied. Although we could not determine the precise topological phase diagram for $N_x=N_y=300$ as we did for $N_x=N_y=20$ in Figs. \ref{fig4} and \ref{fig5}, we can confirm that Fig. \ref{fig3}(e) with $\mu-E_D=-0.33$ and $E_Z=0.3\Delta_0$ does not feature MZMs by increasing the resolution of the LDOS calculated with the KPM. Similar to Fig. \ref{fig2}(g-i), we only choose the sites with the highest LDOS at $E = 0$ and
$N = 2000$ for calculations with high resolution. In Fig. \ref{figF2}(a, c), there are no MZMs
since the LDOS does not peak at $E=0$. For $B_{/\hspace{-1mm}/}$ along $[100]$, as $M_T$ is broken, the small peak at $E = 0$ does not imply existence of MZMs, it originates from bulk states with energy much lower than the resolution.

\begin{figure}[H]
	\centering
	\includegraphics[width=0.98\columnwidth,center]{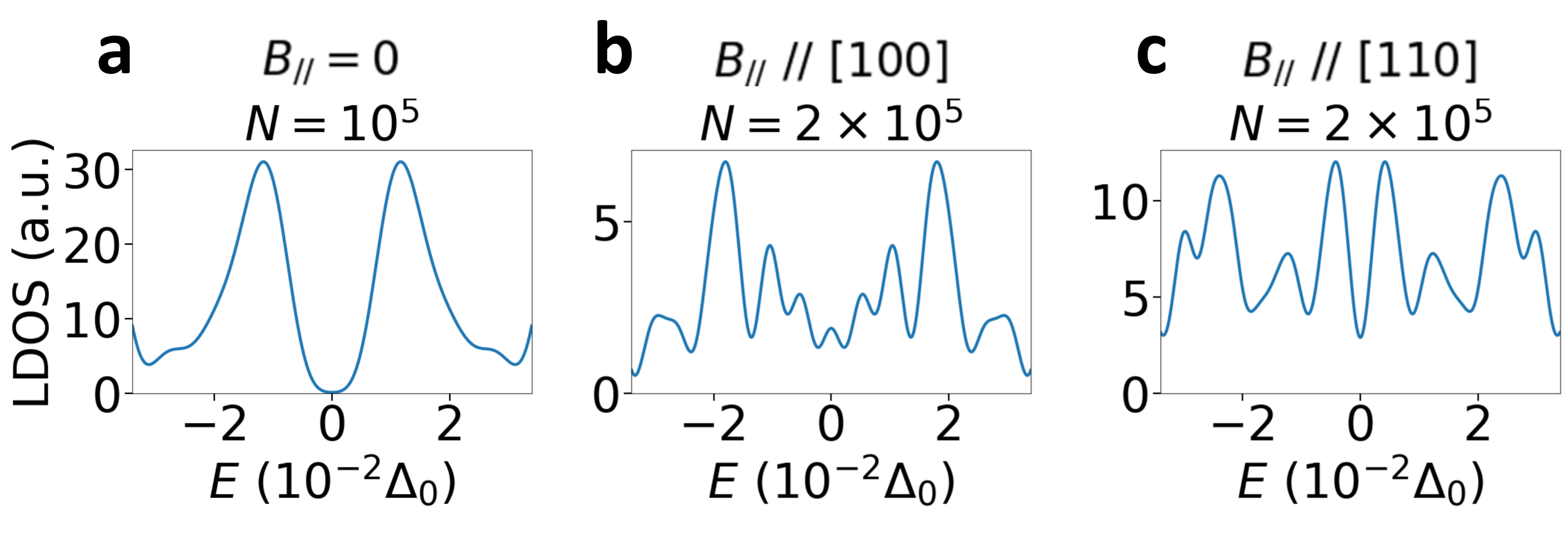}
	\caption{\label{figF2} LDOS at high resolution for $\mu-E_D=-0.33$. In vertical magnetic field (a), in tilted magnetic fields with  $B_{/\hspace{-1mm}/}$ along $[100]$ (b), and along $[110]$ (c). We choose the sites with the highest LDOS at $E=0$ for $N=2000$. For (b, c) the Zeeman energy is $E_Z=0.3\Delta_0$.}
\end{figure}

\begin{figure}[H]
	\centering
	\includegraphics[width=0.98\columnwidth,center]{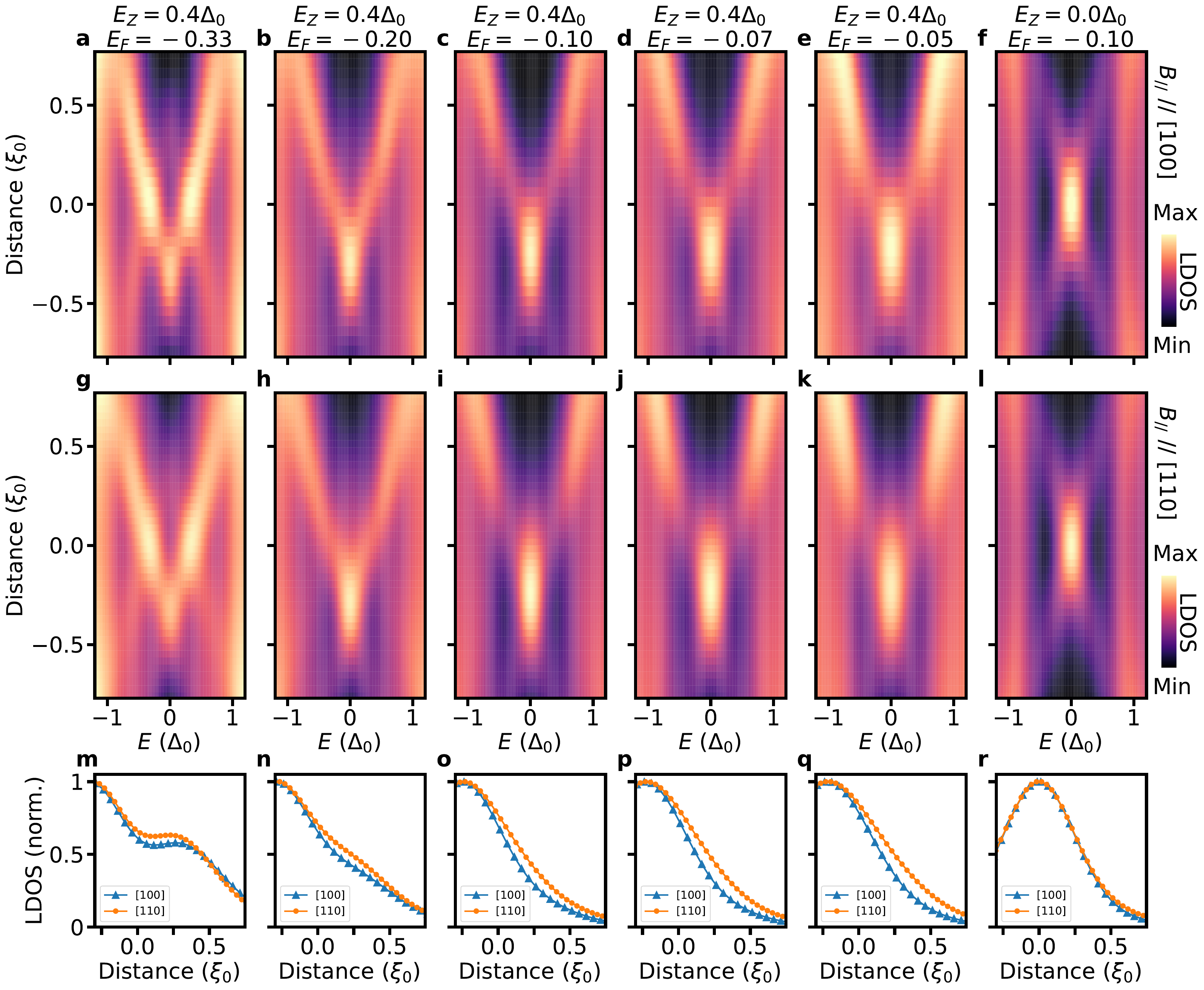}
	\caption{\label{figF3} Vortex LDOS for the (001) surface of SnTe in tilted magnetic fields at various Fermi energy $E_F$ ($E_F=\mu-E_D$) and $E_Z=0.4\Delta_0$. The number of moments is $N=2000$. The bottom row shows the LDOS at $E=0$ for either scanning direction which is also the direction of  $B_{/\hspace{-1mm}/}$. The rightmost column shows the LDOS in vertical magnetic field for comparisons, there is a slight anisotropy in the LDOS due to the square lattice (r). Other parameters are the same as in Fig. \ref{fig2}.}
\end{figure}

In Fig. \ref{figF3} we show the vortex LDOS with different $E_F$ for $B_{/\hspace{-1mm}/}$ along $[100]$ and $[110]$ with fixed Zeeman energy $E_Z=0.4\Delta_0$. We define $E_F=\mu-E_D$ to shorten the notations in the figure. At $E_F=-0.33$ far below the Dirac points, the LDOS have a similar V-shape in either directions since they are due to the nearly isotropic bulk states. Since the vortex states are mainly contributed by bulk states that decay slowly beneath the surface, and the thickness of our lattice $(N_z=50)$ is comparable to the vortex size, the vortex LDOS on the top surface has contribution from the bottom surface which appears as an upside-down V shape in Fig. \ref{figF3} (a, g). As $E_F$ increases (closer to the Dirac points), the magnetic response becomes more anisotropic. The ZBP elongates for $B_{/\hspace{-1mm}/}$ along $[110]$ and deviates from the V shape when $B_{/\hspace{-1mm}/}$ is along $[100]$. And the LDOS does not depend sensitively on the phase transition that destroy the MZMs, the change in the magnetic response originates from competition between the MZMs and the bulk states as explained in Sec. \ref{sec4b}.

	\section{Symmetry operators for the tight-binding model}\label{appendixA}
	\setcounter{figure}{0} 
	The symmetry operators for the BdG Hamiltonian for a vortex line (eq. 2 in the main text) can be decomposed into 5 parts
	\begin{equation}
		O_{\text{vortex}}=O_{\text{ph}}\otimes O_{x,y}\otimes O_z \otimes O_{\text{orb}} \otimes O_{\text{spin}}.
	\end{equation}
	$O_{\text{ph}}$ is in the Nambu basis. $O_{xy}$ and $O_{z}$ are for the sites in the supercell for the vortex line, $z=0,1$ since there are 2 sublattices. $O_{\text{orb}}$ is for orbitals $p_x$, $p_y$ and $p_z$. $O_{\text{spin}}$ is for the spin basis.
	
	The particle-hole symmetry operator is
	\begin{equation}
		C_{\text{ph}}=
		\left(\begin{array}{cc}
			0 & K \\
			K & 0
		\end{array}\right),
	\end{equation}
	where $K$ is the complex conjugation operator. The Hamiltonian satisfies $CH_{\text{BdG}}(k_z)C^{-1}=-H_{\text{BdG}}(-k_z)$ with $C^2=+1$.
	
	The magnetic mirror symmetry $M_T=TM_{1\bar{1}0}$ is written as
	\begin{equation}
		M_T=
		\left(\begin{array}{cc}
			e^{i\pi/4}i\sigma_ym_{1\bar{1}0}^{*}K & 0 \\
			0 & e^{-i\pi/4}i\sigma_ym_{1\bar{1}0}K
		\end{array}\right),
	\end{equation}
	where $\sigma_y$ is a Pauli matrix in the spin basis. The phase factors $e^{\pm i \pi/4}$ are introduced to cancel the phases generated when rotating the vortex \cite{qin2019quasi}. And $m_{1\bar{1}0}$ is the mirror symmetry operator for the band Hamiltonian
	\begin{equation}
		m_{1\bar{1}0}=m_{1\bar{1}0}^{(xy)}\otimes\mathbb{I}_2\otimes\left(\begin{array}{ccc}
			0 & 1 & 0 \\
			1 & 0 & 0 \\
			0 & 0 & 1
		\end{array}\right)\otimes\frac{i}{\sqrt{2}}\left(
		\sigma_x+\sigma_y
		\right),
	\end{equation}
	where $m_{1\bar{1}0}^{(xy)}$ transforms $(x,y)$ to $(y,x)$ and $\mathbb{I}_2$ is an identity in the $z$ basis. And the origin $(0,0)$ is the center of the lattice.
	
	The Hamiltonian satisfies $M_TH_{\text{BdG}}(k_z)M_T^{-1}=H_{BdG}(-k_z)$ with $M_T^2=+1$, it acts as an effective time-reversal symmetry. A chiral symmetry operator can be defined by $\Gamma=M_TC$, The Hamiltonian satisfies $\Gamma H_{\text{BdG}}(k_z)\Gamma^{-1}=-H_{\text{BdG}}(k_z)$ with $\Gamma^2=+1$ \cite{schnyder2008classification,chiu2016classification,xiong2017anisotropic}. In this case the 1D vortex line in SnTe belongs to Altland-Zirnbauer class BDI with a topological invariant of integer values \cite{tewari2012topological, schnyder2008classification}.
	
	For even $N_x$ and $N_y$, the system preserves nonsymmorphic symmetries that interchanges the sublattices \cite{kobayashi2023crystal}. The magnetic glide symmetry $G_T=TG_y$ is written as
	\begin{equation}
		G_T=
		\left(\begin{array}{cc}
			i\sigma_y g_{y}^{*}(k_z)K & 0 \\
			0 & i\sigma_yg_y(-k_z)K
		\end{array}\right),
	\end{equation}
	\begin{equation}
		g_{y}=g_{y}^{(xy)}\otimes\left(\begin{array}{cc}
			0 & e^{-ik_za/2} \\
			e^{ik_za/2} & 0
		\end{array}\right)\otimes\left(\begin{array}{ccc}
			1 & 0 & 0 \\
			0 & -1 & 0 \\
			0 & 0 & 1
		\end{array}\right)\otimes(-i\sigma_y)
	\end{equation}
	and $g_{y}^{(xy)}$ transforms $(x,y)$ to $(x,-y)$. And $G_T^2=-1$ at $k_z=-\pi/a$ guarantees a Kramers degeneracy for the vortex line.
	
	And the screw symmetry operator $S_{4z}$ is written as
	\begin{equation}
		S_{4z}=
		\left(\begin{array}{cc}
			e^{i\pi/4}s_{4z}(k_z) & 0 \\
			0 & e^{-i\pi/4}s_{4z}^*(-k_z)K
		\end{array}\right),
	\end{equation}
	\begin{equation}
		\begin{split}
			s_{4z}=s_{4z}^{(xy)}\otimes\left(\begin{array}{cc}
				0 & e^{ik_za/2} \\
				e^{-ik_za/2} & 0
			\end{array}\right)\otimes&\left(\begin{array}{ccc}
				0 & -1 & 0 \\
				1 & 0 & 0 \\
				0 & 0 & 1
			\end{array}\right)\\
			\otimes&\frac{1}{\sqrt{2}}(\sigma_0-i\sigma_z)
		\end{split}
	\end{equation}
	and $s_{4z}^{(xy)}$ transforms $(x,y)$ to $(-y,x)$.

	\section{Recursive Green's Functions}\label{appendixB}
	\setcounter{figure}{0}
	We use recursive Green's functions \cite{sancho1985highly} to calculate the number of zero-energy states near the surface. The LDOS $d^i(E)$ is defined in Eq. 3 of the main text.
	We calculate the number of zero-energy states by integrating the LDOS using $\eta$ smaller than the lowest nonzero eigenenergy. We checked that for our choice of parameters we can use $\eta=10^{-6}$, and $d^i(E)$ can be described by a single Lorentz function.
	\begin{equation}\label{eq_b_dos2}
		d^i(E)\approx\frac{1}{\pi}\sum_{E_k=0}\langle i | k \rangle \langle k | i \rangle \frac{\eta}{(E-E_k)^2+\eta^2}.
	\end{equation}
	The energy integration can be replaced by $\pi\eta d^i(E=0)$.
	
\begin{figure}[H]
		\centering
		\includegraphics[width=0.8\columnwidth,center]{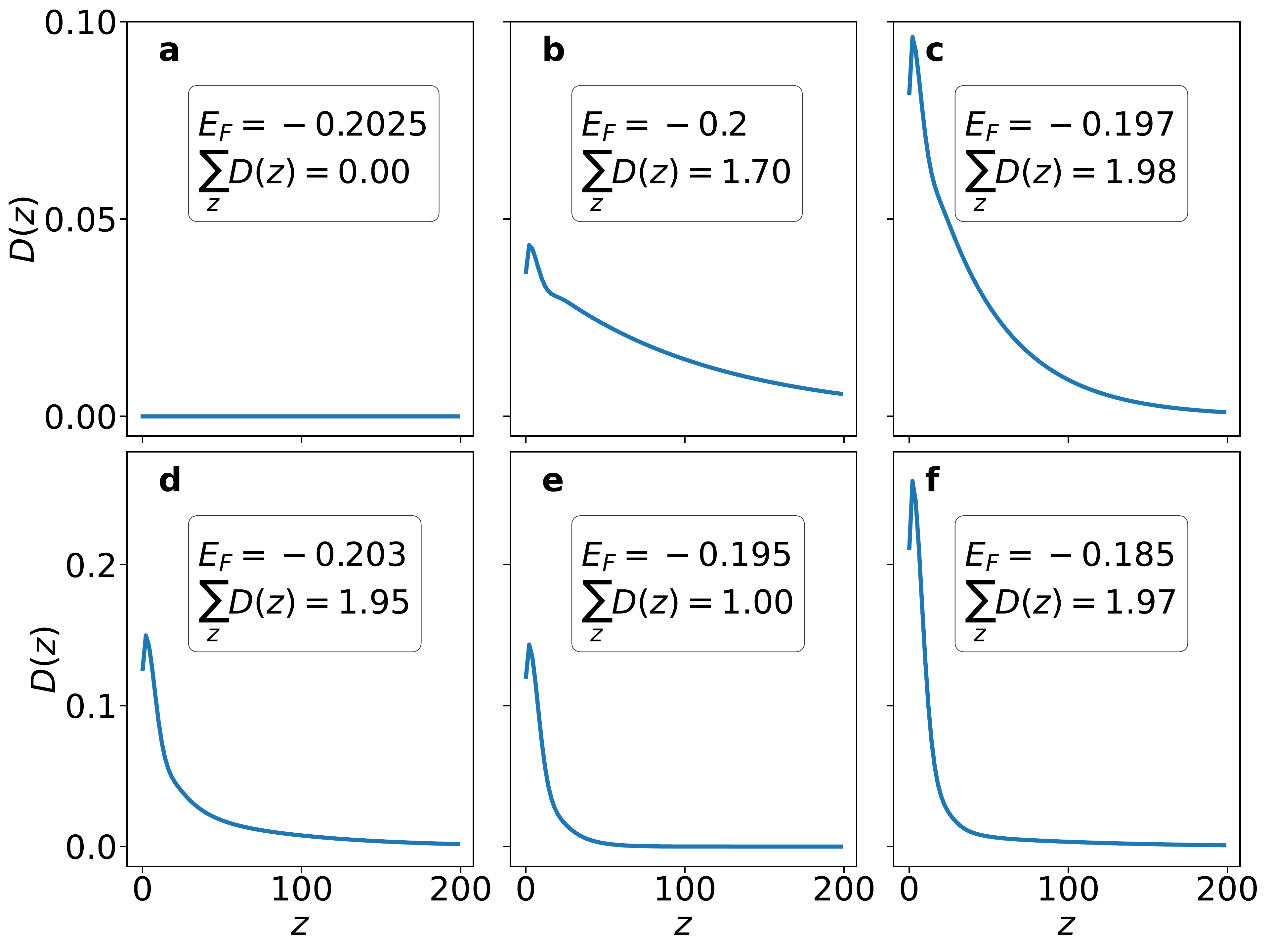}
		\caption{\label{figB1} Distribution of the MZMs $D(z)$ for the top $n_z=200$ layers. Zeeman energy $E_Z=0$ in (a-c) and $E_Z=0.015$ in (d-f). $E_F=\mu-E_D$ as defined in the main text.}
	\end{figure}

	Since the Hamiltonian has no $z$ dependence, the surface Green's functions can be obtained recursively \cite{sancho1985highly}. And we define the LDOS for each layer at zero energy as $D(z)$. We iterate the surface Green's functions to obtain the LDOS for $n_z$ layers below the surface. Then the number of MZMs can be approximated by
	\begin{equation}\label{eq_b5}
		N=\sum_{z=1}^{n_z}D(z),
	\end{equation}
	the results are shown in Fig. 4(b) in the main text using $n_z=100,200$. In Fig. \ref{figB1} we show $D(z)$ for some values of $(\mu,E_Z)$. Near the phase transition points the MZMs penetrate too deep inside the bulk \cite{li2014majorana,kawakami2015evolution}, and $N$ deviates from an integer value [e.g., Fig. 4(b) in the main text].
	
	\section{Winding number}\label{appendixC}
	\setcounter{figure}{0}
	The number of MZMs is equal to the winding number
	\begin{equation}\label{eq_c1}
		\nu=\frac{1}{2\pi}\int^{2\pi}_0d\theta(k_z),
	\end{equation}
	where $\theta(k_z)$ is the angle of the $\pmb{d}$ vector in Nambu space.
	For a vortex line (eq. 2 in the main text) the winding number is generalized to include other degrees of freedom (sites, spins, orbitals, etc.), which is defined as \cite{tewari2012topological,sato2016majorana,kobayashi2020double}
	\begin{equation}\label{eq_c2}
		\nu=\frac{1}{4\pi i}\int^{2\pi}_0dk_z \text{Tr}\Bigg[ \Gamma \frac{H_{BdG}(k_z)}{dk_z} H^{-1}_{BdG}(k_z) \Bigg].
	\end{equation}
	We show the integrands for some values of $(\mu,E_Z)$ in Fig. \ref{figC1}. Previous references for vortex phase transitions only calculated the crossing of vortex line to deduce the change of the number of MZMs, which leads to ambiguous results when the system preserves a chiral symmetry that supports multiple MZMs. In Fig. \ref{figC1} (d-f), we explicitly show that $\nu$ changes as $2 \rightarrow 1 \rightarrow 2$, and the result is consistent with the number of MZMs at each end obtained in the last section.
	
	\begin{figure}[H]
		\centering
		\includegraphics[width=0.8\columnwidth,center]{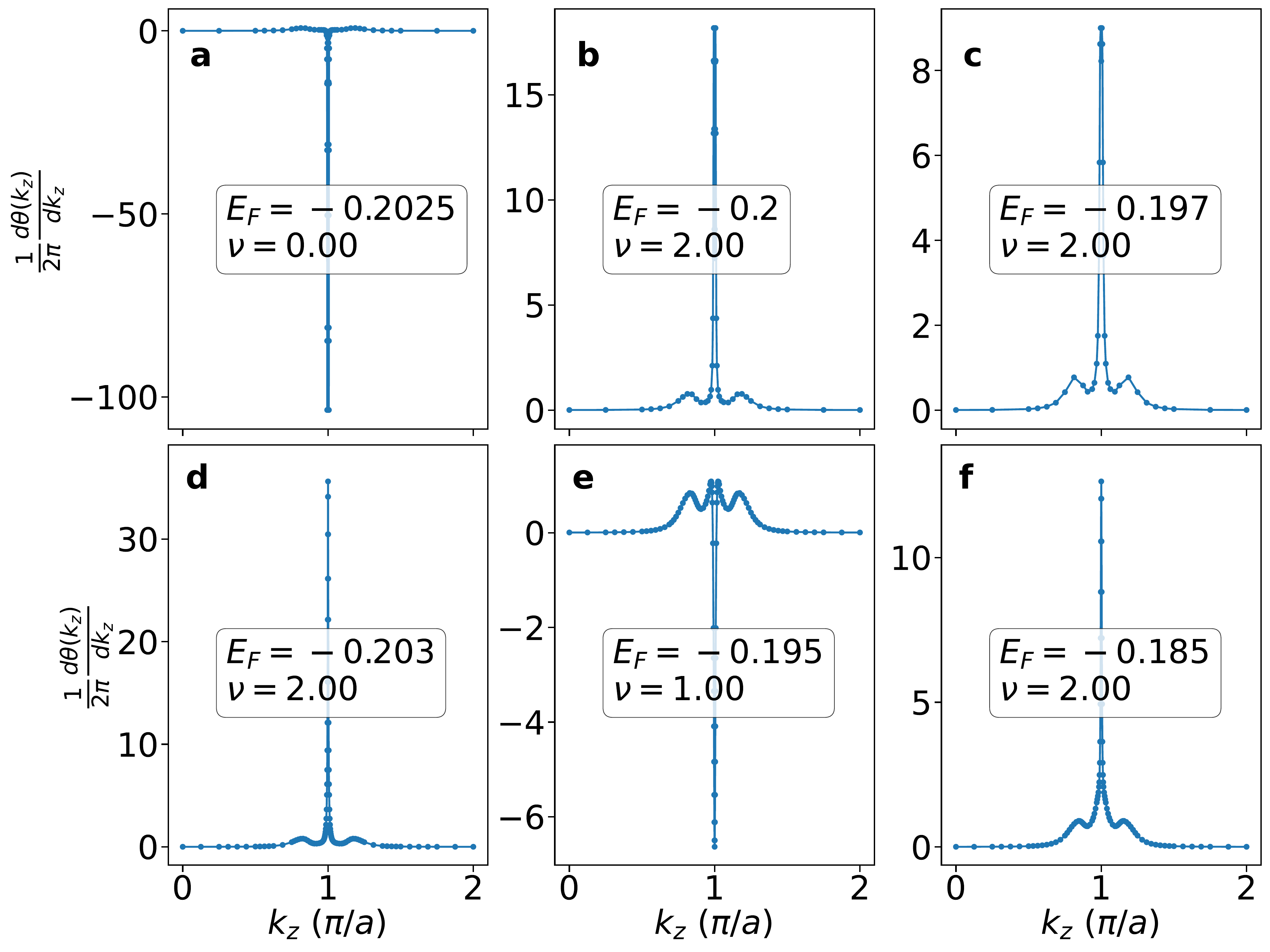}
		\caption{\label{figC1} The integrands of the winding number. Zeeman energy $E_Z=0$ in (a-c) and $E_Z=0.015$ in (d-f).}
	\end{figure}
	
	\section{Vortex phase transitions for odd $N_x=N_y$}\label{appendixD}
	\setcounter{figure}{0}
	In the main text we have shown that the single MZM phase may appears between vortex phase transitions if $G_T$ is broken, and confirmed it using an in-plane Zeeman field that breaks $G_T$. In Fig. \ref{figD1} we show $\nu=1$ can also be realized in system with odd $N_x=N_y$ that breaks $G_T$. The calculations are the same as those for Fig. 4 in the main text, except that in Fig. \ref{figD1} (d-g) we show the $C_{4z}$ eigenvalues since $S_{4z}$ is broken.
	\begin{figure}[H]
		\centering
		\includegraphics[width=0.98\columnwidth,center]{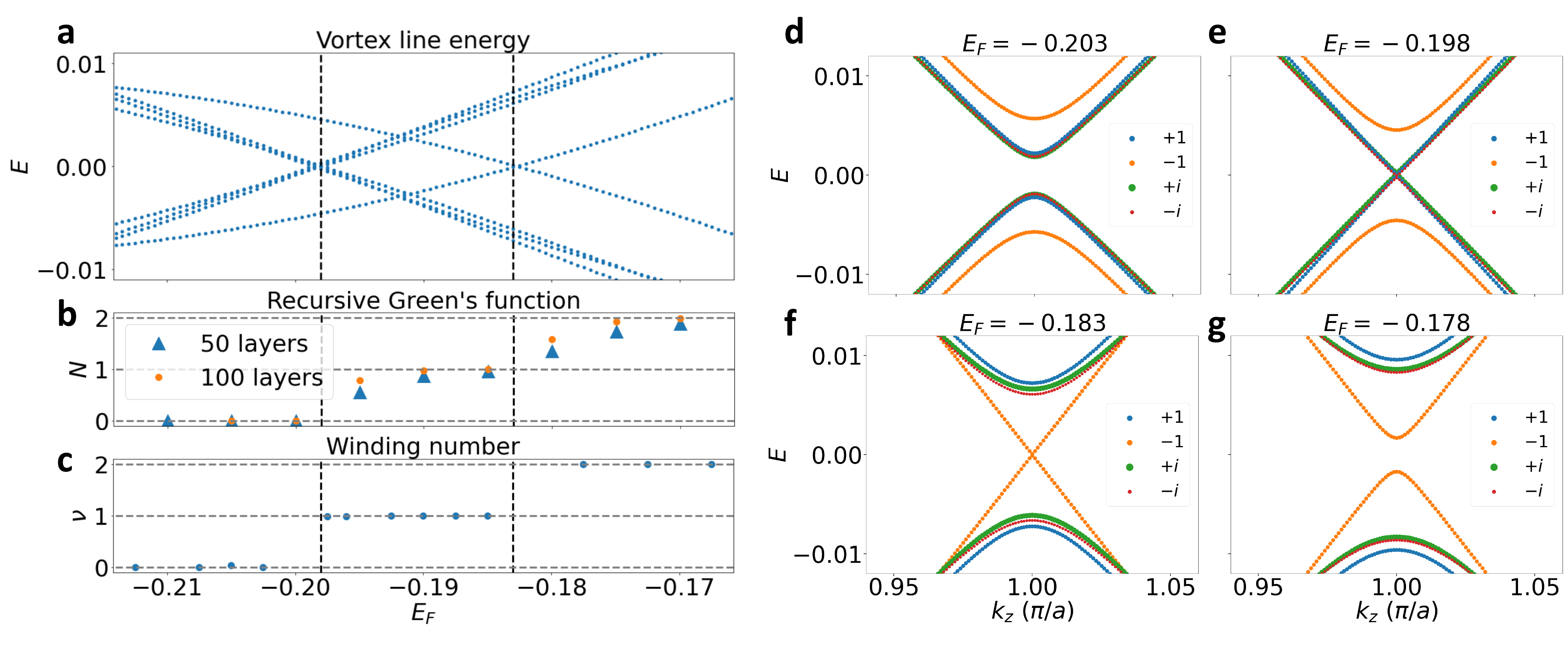}
		\caption{\label{figD1} Vortex phase transition in SnTe under an out-of-plane magnetic field. (a) The vortex line energy at $k_z=\pi/a$, it can be non-degenerate since $G_T$ is broken. The number of MZMs are calculated using recursive Green's functions (b) and winding number (c). The black dashed lines denote crossings at $k_z=\pi/a$. (d-g) Vortex line energy for different $k_z$, the colors denote the $C_{4z}$ eigenvalues. We use $\Delta_0=0.1$, $\xi_0=1$ for the vortex and $N_x=N_y=21$ for the lattice.}
	\end{figure}
	
\end{appendices}
\bibliography{sn-article}


\begin{thebibliography}{102}
\ifx \bisbn   \undefined \def \bisbn  #1{ISBN #1}\fi
\ifx \binits  \undefined \def \binits#1{#1}\fi
\ifx \bauthor  \undefined \def \bauthor#1{#1}\fi
\ifx \batitle  \undefined \def \batitle#1{#1}\fi
\ifx \bjtitle  \undefined \def \bjtitle#1{#1}\fi
\ifx \bvolume  \undefined \def \bvolume#1{\textbf{#1}}\fi
\ifx \byear  \undefined \def \byear#1{#1}\fi
\ifx \bissue  \undefined \def \bissue#1{#1}\fi
\ifx \bfpage  \undefined \def \bfpage#1{#1}\fi
\ifx \blpage  \undefined \def \blpage #1{#1}\fi
\ifx \burl  \undefined \def \burl#1{\textsf{#1}}\fi
\ifx \doiurl  \undefined \def \doiurl#1{\url{https://doi.org/#1}}\fi
\ifx \betal  \undefined \def \betal{\textit{et al.}}\fi
\ifx \binstitute  \undefined \def \binstitute#1{#1}\fi
\ifx \binstitutionaled  \undefined \def \binstitutionaled#1{#1}\fi
\ifx \bctitle  \undefined \def \bctitle#1{#1}\fi
\ifx \beditor  \undefined \def \beditor#1{#1}\fi
\ifx \bpublisher  \undefined \def \bpublisher#1{#1}\fi
\ifx \bbtitle  \undefined \def \bbtitle#1{#1}\fi
\ifx \bedition  \undefined \def \bedition#1{#1}\fi
\ifx \bseriesno  \undefined \def \bseriesno#1{#1}\fi
\ifx \blocation  \undefined \def \blocation#1{#1}\fi
\ifx \bsertitle  \undefined \def \bsertitle#1{#1}\fi
\ifx \bsnm \undefined \def \bsnm#1{#1}\fi
\ifx \bsuffix \undefined \def \bsuffix#1{#1}\fi
\ifx \bparticle \undefined \def \bparticle#1{#1}\fi
\ifx \barticle \undefined \def \barticle#1{#1}\fi
\bibcommenthead
\ifx \bconfdate \undefined \def \bconfdate #1{#1}\fi
\ifx \botherref \undefined \def \botherref #1{#1}\fi
\ifx \url \undefined \def \url#1{\textsf{#1}}\fi
\ifx \bchapter \undefined \def \bchapter#1{#1}\fi
\ifx \bbook \undefined \def \bbook#1{#1}\fi
\ifx \bcomment \undefined \def \bcomment#1{#1}\fi
\ifx \oauthor \undefined \def \oauthor#1{#1}\fi
\ifx \citeauthoryear \undefined \def \citeauthoryear#1{#1}\fi
\ifx \endbibitem  \undefined \def \endbibitem {}\fi
\ifx \bconflocation  \undefined \def \bconflocation#1{#1}\fi
\ifx \arxivurl  \undefined \def \arxivurl#1{\textsf{#1}}\fi
\csname PreBibitemsHook\endcsname

%%% 1
\bibitem[\protect\citeauthoryear{Kitaev}{2003}]{kitaev2003fault}
\begin{barticle}
\bauthor{\bsnm{Kitaev}, \binits{A.Y.}}:
\batitle{Fault-tolerant quantum computation by anyons}.
\bjtitle{Annals of Physics}
\bvolume{303}(\bissue{1}),
\bfpage{2}--\blpage{30}
(\byear{2003})
\end{barticle}
\endbibitem

%%% 2
\bibitem[\protect\citeauthoryear{Alicea}{2012}]{alicea2012new}
\begin{barticle}
\bauthor{\bsnm{Alicea}, \binits{J.}}:
\batitle{New directions in the pursuit of {M}ajorana fermions in solid state
  systems}.
\bjtitle{Reports on Progress in Physics}
\bvolume{75}(\bissue{7}),
\bfpage{076501}
(\byear{2012})
\end{barticle}
\endbibitem

%%% 3
\bibitem[\protect\citeauthoryear{Sarma et~al.}{2015}]{sarma2015majorana}
\begin{barticle}
\bauthor{\bsnm{Sarma}, \binits{S.D.}},
\bauthor{\bsnm{Freedman}, \binits{M.}},
\bauthor{\bsnm{Nayak}, \binits{C.}}:
\batitle{Majorana zero modes and topological quantum computation}.
\bjtitle{npj Quantum Information}
\bvolume{1}(\bissue{1}),
\bfpage{1}--\blpage{13}
(\byear{2015})
\end{barticle}
\endbibitem

%%% 4
\bibitem[\protect\citeauthoryear{Aasen et~al.}{2016}]{aasen2016milestones}
\begin{barticle}
\bauthor{\bsnm{Aasen}, \binits{D.}},
\bauthor{\bsnm{Hell}, \binits{M.}},
\bauthor{\bsnm{Mishmash}, \binits{R.V.}},
\bauthor{\bsnm{Higginbotham}, \binits{A.}},
\bauthor{\bsnm{Danon}, \binits{J.}},
\bauthor{\bsnm{Leijnse}, \binits{M.}},
\bauthor{\bsnm{Jespersen}, \binits{T.S.}},
\bauthor{\bsnm{Folk}, \binits{J.A.}},
\bauthor{\bsnm{Marcus}, \binits{C.M.}},
\bauthor{\bsnm{Flensberg}, \binits{K.}}, \betal:
\batitle{Milestones toward {Majorana-based} quantum computing}.
\bjtitle{Physical Review X}
\bvolume{6}(\bissue{3}),
\bfpage{031016}
(\byear{2016})
\end{barticle}
\endbibitem

%%% 5
\bibitem[\protect\citeauthoryear{Lutchyn et~al.}{2018}]{lutchyn2018majorana}
\begin{barticle}
\bauthor{\bsnm{Lutchyn}, \binits{R.M.}},
\bauthor{\bsnm{Bakkers}, \binits{E.P.}},
\bauthor{\bsnm{Kouwenhoven}, \binits{L.P.}},
\bauthor{\bsnm{Krogstrup}, \binits{P.}},
\bauthor{\bsnm{Marcus}, \binits{C.M.}},
\bauthor{\bsnm{Oreg}, \binits{Y.}}:
\batitle{Majorana zero modes in superconductor--semiconductor
  heterostructures}.
\bjtitle{Nature Reviews Materials}
\bvolume{3}(\bissue{5}),
\bfpage{52}--\blpage{68}
(\byear{2018})
\end{barticle}
\endbibitem

%%% 6
\bibitem[\protect\citeauthoryear{Beenakker}{2020}]{beenakker2020search}
\begin{botherref}
\oauthor{\bsnm{Beenakker}, \binits{C.}}:
Search for non-{A}belian {M}ajorana braiding statistics in superconductors.
SciPost Physics Lecture Notes,
015
(2020)
\end{botherref}
\endbibitem

%%% 7
\bibitem[\protect\citeauthoryear{Awoga et~al.}{2024}]{awoga2024controlling}
\begin{barticle}
\bauthor{\bsnm{Awoga}, \binits{O.A.}},
\bauthor{\bsnm{Ioannidis}, \binits{I.}},
\bauthor{\bsnm{Mishra}, \binits{A.}},
\bauthor{\bsnm{Leijnse}, \binits{M.}},
\bauthor{\bsnm{Trif}, \binits{M.}},
\bauthor{\bsnm{Posske}, \binits{T.}}:
\batitle{Controlling {M}ajorana hybridization in magnetic chain-superconductor
  systems}.
\bjtitle{Physical Review Research}
\bvolume{6}(\bissue{3}),
\bfpage{033154}
(\byear{2024})
\end{barticle}
\endbibitem

%%% 8
\bibitem[\protect\citeauthoryear{Mourik et~al.}{2012}]{mourik2012signatures}
\begin{barticle}
\bauthor{\bsnm{Mourik}, \binits{V.}},
\bauthor{\bsnm{Zuo}, \binits{K.}},
\bauthor{\bsnm{Frolov}, \binits{S.M.}},
\bauthor{\bsnm{Plissard}, \binits{S.}},
\bauthor{\bsnm{Bakkers}, \binits{E.P.}},
\bauthor{\bsnm{Kouwenhoven}, \binits{L.P.}}:
\batitle{Signatures of {M}ajorana fermions in hybrid
  superconductor-semiconductor nanowire devices}.
\bjtitle{Science}
\bvolume{336}(\bissue{6084}),
\bfpage{1003}--\blpage{1007}
(\byear{2012})
\end{barticle}
\endbibitem

%%% 9
\bibitem[\protect\citeauthoryear{Das et~al.}{2012}]{das2012zero}
\begin{barticle}
\bauthor{\bsnm{Das}, \binits{A.}},
\bauthor{\bsnm{Ronen}, \binits{Y.}},
\bauthor{\bsnm{Most}, \binits{Y.}},
\bauthor{\bsnm{Oreg}, \binits{Y.}},
\bauthor{\bsnm{Heiblum}, \binits{M.}},
\bauthor{\bsnm{Shtrikman}, \binits{H.}}:
\batitle{Zero-bias peaks and splitting in an {A}l--{I}n{A}s nanowire
  topological superconductor as a signature of {M}ajorana fermions}.
\bjtitle{Nature Physics}
\bvolume{8}(\bissue{12}),
\bfpage{887}--\blpage{895}
(\byear{2012})
\end{barticle}
\endbibitem

%%% 10
\bibitem[\protect\citeauthoryear{Nadj-Perge et~al.}{2014}]{nadj2014observation}
\begin{barticle}
\bauthor{\bsnm{Nadj-Perge}, \binits{S.}},
\bauthor{\bsnm{Drozdov}, \binits{I.K.}},
\bauthor{\bsnm{Li}, \binits{J.}},
\bauthor{\bsnm{Chen}, \binits{H.}},
\bauthor{\bsnm{Jeon}, \binits{S.}},
\bauthor{\bsnm{Seo}, \binits{J.}},
\bauthor{\bsnm{MacDonald}, \binits{A.H.}},
\bauthor{\bsnm{Bernevig}, \binits{B.A.}},
\bauthor{\bsnm{Yazdani}, \binits{A.}}:
\batitle{Observation of {M}ajorana fermions in ferromagnetic atomic chains on a
  superconductor}.
\bjtitle{Science}
\bvolume{346}(\bissue{6209}),
\bfpage{602}--\blpage{607}
(\byear{2014})
\end{barticle}
\endbibitem

%%% 11
\bibitem[\protect\citeauthoryear{Xu et~al.}{2015}]{xu2015experimental}
\begin{barticle}
\bauthor{\bsnm{Xu}, \binits{J.-P.}},
\bauthor{\bsnm{Wang}, \binits{M.-X.}},
\bauthor{\bsnm{Liu}, \binits{Z.L.}},
\bauthor{\bsnm{Ge}, \binits{J.-F.}},
\bauthor{\bsnm{Yang}, \binits{X.}},
\bauthor{\bsnm{Liu}, \binits{C.}},
\bauthor{\bsnm{Xu}, \binits{Z.A.}},
\bauthor{\bsnm{Guan}, \binits{D.}},
\bauthor{\bsnm{Gao}, \binits{C.L.}},
\bauthor{\bsnm{Qian}, \binits{D.}}, \betal:
\batitle{Experimental detection of a majorana mode in the core of a magnetic
  vortex inside a topological insulator-superconductor
  {Bi}$_2${Te}$_3$/{Nb}{Se}$_2$ heterostructure}.
\bjtitle{Physical Review Letters}
\bvolume{114}(\bissue{1}),
\bfpage{017001}
(\byear{2015})
\end{barticle}
\endbibitem

%%% 12
\bibitem[\protect\citeauthoryear{Yin et~al.}{2015}]{yin2015observation}
\begin{barticle}
\bauthor{\bsnm{Yin}, \binits{J.-X.}},
\bauthor{\bsnm{Wu}, \binits{Z.}},
\bauthor{\bsnm{Wang}, \binits{J.}},
\bauthor{\bsnm{Ye}, \binits{Z.}},
\bauthor{\bsnm{Gong}, \binits{J.}},
\bauthor{\bsnm{Hou}, \binits{X.}},
\bauthor{\bsnm{Shan}, \binits{L.}},
\bauthor{\bsnm{Li}, \binits{A.}},
\bauthor{\bsnm{Liang}, \binits{X.}},
\bauthor{\bsnm{Wu}, \binits{X.}}, \betal:
\batitle{Observation of a robust zero-energy bound state in iron-based
  superconductor {Fe}({Te}, {Se})}.
\bjtitle{Nature Physics}
\bvolume{11}(\bissue{7}),
\bfpage{543}--\blpage{546}
(\byear{2015})
\end{barticle}
\endbibitem

%%% 13
\bibitem[\protect\citeauthoryear{Ruby et~al.}{2015}]{ruby2015end}
\begin{barticle}
\bauthor{\bsnm{Ruby}, \binits{M.}},
\bauthor{\bsnm{Pientka}, \binits{F.}},
\bauthor{\bsnm{Peng}, \binits{Y.}},
\bauthor{\bsnm{Von~Oppen}, \binits{F.}},
\bauthor{\bsnm{Heinrich}, \binits{B.W.}},
\bauthor{\bsnm{Franke}, \binits{K.J.}}:
\batitle{End states and subgap structure in proximity-coupled chains of
  magnetic adatoms}.
\bjtitle{Physical Review Letters}
\bvolume{115}(\bissue{19}),
\bfpage{197204}
(\byear{2015})
\end{barticle}
\endbibitem

%%% 14
\bibitem[\protect\citeauthoryear{M{\'e}nard et~al.}{2017}]{menard2017two}
\begin{barticle}
\bauthor{\bsnm{M{\'e}nard}, \binits{G.C.}},
\bauthor{\bsnm{Guissart}, \binits{S.}},
\bauthor{\bsnm{Brun}, \binits{C.}},
\bauthor{\bsnm{Leriche}, \binits{R.T.}},
\bauthor{\bsnm{Trif}, \binits{M.}},
\bauthor{\bsnm{Debontridder}, \binits{F.}},
\bauthor{\bsnm{Demaille}, \binits{D.}},
\bauthor{\bsnm{Roditchev}, \binits{D.}},
\bauthor{\bsnm{Simon}, \binits{P.}},
\bauthor{\bsnm{Cren}, \binits{T.}}:
\batitle{Two-dimensional topological superconductivity in {P}b/{C}o/{S}i
  (111)}.
\bjtitle{Nature Communications}
\bvolume{8}(\bissue{1}),
\bfpage{1}--\blpage{7}
(\byear{2017})
\end{barticle}
\endbibitem

%%% 15
\bibitem[\protect\citeauthoryear{Zhang et~al.}{2018}]{zhang2018observation}
\begin{barticle}
\bauthor{\bsnm{Zhang}, \binits{P.}},
\bauthor{\bsnm{Yaji}, \binits{K.}},
\bauthor{\bsnm{Hashimoto}, \binits{T.}},
\bauthor{\bsnm{Ota}, \binits{Y.}},
\bauthor{\bsnm{Kondo}, \binits{T.}},
\bauthor{\bsnm{Okazaki}, \binits{K.}},
\bauthor{\bsnm{Wang}, \binits{Z.}},
\bauthor{\bsnm{Wen}, \binits{J.}},
\bauthor{\bsnm{Gu}, \binits{G.}},
\bauthor{\bsnm{Ding}, \binits{H.}}, \betal:
\batitle{Observation of topological superconductivity on the surface of an
  iron-based superconductor}.
\bjtitle{Science}
\bvolume{360}(\bissue{6385}),
\bfpage{182}--\blpage{186}
(\byear{2018})
\end{barticle}
\endbibitem

%%% 16
\bibitem[\protect\citeauthoryear{Wang et~al.}{2018}]{wang2018evidence}
\begin{barticle}
\bauthor{\bsnm{Wang}, \binits{D.}},
\bauthor{\bsnm{Kong}, \binits{L.}},
\bauthor{\bsnm{Fan}, \binits{P.}},
\bauthor{\bsnm{Chen}, \binits{H.}},
\bauthor{\bsnm{Zhu}, \binits{S.}},
\bauthor{\bsnm{Liu}, \binits{W.}},
\bauthor{\bsnm{Cao}, \binits{L.}},
\bauthor{\bsnm{Sun}, \binits{Y.}},
\bauthor{\bsnm{Du}, \binits{S.}},
\bauthor{\bsnm{Schneeloch}, \binits{J.}}, \betal:
\batitle{Evidence for {M}ajorana bound states in an iron-based superconductor}.
\bjtitle{Science}
\bvolume{362}(\bissue{6412}),
\bfpage{333}--\blpage{335}
(\byear{2018})
\end{barticle}
\endbibitem

%%% 17
\bibitem[\protect\citeauthoryear{Liu et~al.}{2018}]{liu2018robust}
\begin{barticle}
\bauthor{\bsnm{Liu}, \binits{Q.}},
\bauthor{\bsnm{Chen}, \binits{C.}},
\bauthor{\bsnm{Zhang}, \binits{T.}},
\bauthor{\bsnm{Peng}, \binits{R.}},
\bauthor{\bsnm{Yan}, \binits{Y.-J.}},
\bauthor{\bsnm{Lou}, \binits{X.}},
\bauthor{\bsnm{Huang}, \binits{Y.-L.}},
\bauthor{\bsnm{Tian}, \binits{J.-P.}},
\bauthor{\bsnm{Dong}, \binits{X.-L.}},
\bauthor{\bsnm{Wang}, \binits{G.-W.}}, \betal:
\batitle{Robust and clean {M}ajorana zero mode in the vortex core of
  high-temperature superconductor ({L}i$_{0.84}${F}e$_{0.16}$){O}{H}{F}e{S}e}.
\bjtitle{Physical Review X}
\bvolume{8}(\bissue{4}),
\bfpage{041056}
(\byear{2018})
\end{barticle}
\endbibitem

%%% 18
\bibitem[\protect\citeauthoryear{Machida et~al.}{2019}]{machida2019zero}
\begin{barticle}
\bauthor{\bsnm{Machida}, \binits{T.}},
\bauthor{\bsnm{Sun}, \binits{Y.}},
\bauthor{\bsnm{Pyon}, \binits{S.}},
\bauthor{\bsnm{Takeda}, \binits{S.}},
\bauthor{\bsnm{Kohsaka}, \binits{Y.}},
\bauthor{\bsnm{Hanaguri}, \binits{T.}},
\bauthor{\bsnm{Sasagawa}, \binits{T.}},
\bauthor{\bsnm{Tamegai}, \binits{T.}}:
\batitle{Zero-energy vortex bound state in the superconducting topological
  surface state of {F}e({S}e, {T}e)}.
\bjtitle{Nature Materials}
\bvolume{18}(\bissue{8}),
\bfpage{811}--\blpage{815}
(\byear{2019})
\end{barticle}
\endbibitem

%%% 19
\bibitem[\protect\citeauthoryear{Manna et~al.}{2020}]{manna2020signature}
\begin{barticle}
\bauthor{\bsnm{Manna}, \binits{S.}},
\bauthor{\bsnm{Wei}, \binits{P.}},
\bauthor{\bsnm{Xie}, \binits{Y.}},
\bauthor{\bsnm{Law}, \binits{K.T.}},
\bauthor{\bsnm{Lee}, \binits{P.A.}},
\bauthor{\bsnm{Moodera}, \binits{J.S.}}:
\batitle{Signature of a pair of {M}ajorana zero modes in superconducting gold
  surface states}.
\bjtitle{Proceedings of the National Academy of Sciences}
\bvolume{117}(\bissue{16}),
\bfpage{8775}--\blpage{8782}
(\byear{2020})
\end{barticle}
\endbibitem

%%% 20
\bibitem[\protect\citeauthoryear{Kezilebieke
  et~al.}{2020}]{kezilebieke2020topological}
\begin{barticle}
\bauthor{\bsnm{Kezilebieke}, \binits{S.}},
\bauthor{\bsnm{Huda}, \binits{M.N.}},
\bauthor{\bsnm{Va{\v{n}}o}, \binits{V.}},
\bauthor{\bsnm{Aapro}, \binits{M.}},
\bauthor{\bsnm{Ganguli}, \binits{S.C.}},
\bauthor{\bsnm{Silveira}, \binits{O.J.}},
\bauthor{\bsnm{G{\l}odzik}, \binits{S.}},
\bauthor{\bsnm{Foster}, \binits{A.S.}},
\bauthor{\bsnm{Ojanen}, \binits{T.}},
\bauthor{\bsnm{Liljeroth}, \binits{P.}}:
\batitle{Topological superconductivity in a van der {W}aals heterostructure}.
\bjtitle{Nature}
\bvolume{588}(\bissue{7838}),
\bfpage{424}--\blpage{428}
(\byear{2020})
\end{barticle}
\endbibitem

%%% 21
\bibitem[\protect\citeauthoryear{Fan et~al.}{2021}]{fan2021observation}
\begin{barticle}
\bauthor{\bsnm{Fan}, \binits{P.}},
\bauthor{\bsnm{Yang}, \binits{F.}},
\bauthor{\bsnm{Qian}, \binits{G.}},
\bauthor{\bsnm{Chen}, \binits{H.}},
\bauthor{\bsnm{Zhang}, \binits{Y.-Y.}},
\bauthor{\bsnm{Li}, \binits{G.}},
\bauthor{\bsnm{Huang}, \binits{Z.}},
\bauthor{\bsnm{Xing}, \binits{Y.}},
\bauthor{\bsnm{Kong}, \binits{L.}},
\bauthor{\bsnm{Liu}, \binits{W.}}, \betal:
\batitle{Observation of magnetic adatom-induced majorana vortex and its
  hybridization with field-induced {M}ajorana vortex in an iron-based
  superconductor}.
\bjtitle{Nature Communications}
\bvolume{12}(\bissue{1}),
\bfpage{1348}
(\byear{2021})
\end{barticle}
\endbibitem

%%% 22
\bibitem[\protect\citeauthoryear{Fu and Kane}{2008}]{fu2008superconducting}
\begin{barticle}
\bauthor{\bsnm{Fu}, \binits{L.}},
\bauthor{\bsnm{Kane}, \binits{C.L.}}:
\batitle{Superconducting proximity effect and {M}ajorana fermions at the
  surface of a topological insulator}.
\bjtitle{Physical Review Letters}
\bvolume{100}(\bissue{9}),
\bfpage{096407}
(\byear{2008})
\end{barticle}
\endbibitem

%%% 23
\bibitem[\protect\citeauthoryear{Alicea et~al.}{2011}]{alicea2011non}
\begin{barticle}
\bauthor{\bsnm{Alicea}, \binits{J.}},
\bauthor{\bsnm{Oreg}, \binits{Y.}},
\bauthor{\bsnm{Refael}, \binits{G.}},
\bauthor{\bsnm{Von~Oppen}, \binits{F.}},
\bauthor{\bsnm{Fisher}, \binits{M.P.}}:
\batitle{Non-{A}belian statistics and topological quantum information
  processing in 1d wire networks}.
\bjtitle{Nature Physics}
\bvolume{7}(\bissue{5}),
\bfpage{412}--\blpage{417}
(\byear{2011})
\end{barticle}
\endbibitem

%%% 24
\bibitem[\protect\citeauthoryear{Pikulin et~al.}{2021}]{pikulin2021protocol}
\begin{botherref}
\oauthor{\bsnm{Pikulin}, \binits{D.I.}},
\oauthor{\bsnm{Heck}, \binits{B.}},
\oauthor{\bsnm{Karzig}, \binits{T.}},
\oauthor{\bsnm{Martinez}, \binits{E.A.}},
\oauthor{\bsnm{Nijholt}, \binits{B.}},
\oauthor{\bsnm{Laeven}, \binits{T.}},
\oauthor{\bsnm{Winkler}, \binits{G.W.}},
\oauthor{\bsnm{Watson}, \binits{J.D.}},
\oauthor{\bsnm{Heedt}, \binits{S.}},
\oauthor{\bsnm{Temurhan}, \binits{M.}}, et al.:
Protocol to identify a topological superconducting phase in a three-terminal
  device.
arXiv preprint arXiv:2103.12217
(2021)
\end{botherref}
\endbibitem

%%% 25
\bibitem[\protect\citeauthoryear{Zhou et~al.}{2022}]{zhou2022fusion}
\begin{barticle}
\bauthor{\bsnm{Zhou}, \binits{T.}},
\bauthor{\bsnm{Dartiailh}, \binits{M.C.}},
\bauthor{\bsnm{Sardashti}, \binits{K.}},
\bauthor{\bsnm{Han}, \binits{J.E.}},
\bauthor{\bsnm{Matos-Abiague}, \binits{A.}},
\bauthor{\bsnm{Shabani}, \binits{J.}},
\bauthor{\bsnm{{\v{Z}}uti{\'c}}, \binits{I.}}:
\batitle{Fusion of {M}ajorana bound states with mini-gate control in
  two-dimensional systems}.
\bjtitle{Nature Communications}
\bvolume{13}(\bissue{1}),
\bfpage{1738}
(\byear{2022})
\end{barticle}
\endbibitem

%%% 26
\bibitem[\protect\citeauthoryear{Fang et~al.}{2014}]{CF14}
\begin{barticle}
\bauthor{\bsnm{Fang}, \binits{C.}},
\bauthor{\bsnm{Gilbert}, \binits{M.J.}},
\bauthor{\bsnm{Bernevig}, \binits{B.A.}}:
\batitle{New class of topological superconductors protected by magnetic group
  symmetries}.
\bjtitle{Physical Review Letters}
\bvolume{112},
\bfpage{106401}
(\byear{2014})
\doiurl{10.1103/PhysRevLett.112.106401}
\end{barticle}
\endbibitem

%%% 27
\bibitem[\protect\citeauthoryear{Liu et~al.}{2014}]{liu2014demonstrating}
\begin{barticle}
\bauthor{\bsnm{Liu}, \binits{X.-J.}},
\bauthor{\bsnm{He}, \binits{J.J.}},
\bauthor{\bsnm{Law}, \binits{K.T.}}:
\batitle{Demonstrating lattice symmetry protection in topological crystalline
  superconductors}.
\bjtitle{Physical Review B}
\bvolume{90}(\bissue{23}),
\bfpage{235141}
(\byear{2014})
\end{barticle}
\endbibitem

%%% 28
\bibitem[\protect\citeauthoryear{Hsieh et~al.}{2012}]{hsieh2012topological}
\begin{barticle}
\bauthor{\bsnm{Hsieh}, \binits{T.H.}},
\bauthor{\bsnm{Lin}, \binits{H.}},
\bauthor{\bsnm{Liu}, \binits{J.}},
\bauthor{\bsnm{Duan}, \binits{W.}},
\bauthor{\bsnm{Bansil}, \binits{A.}},
\bauthor{\bsnm{Fu}, \binits{L.}}:
\batitle{Topological crystalline insulators in the {S}n{T}e material class}.
\bjtitle{Nature Communications}
\bvolume{3}(\bissue{1}),
\bfpage{1}--\blpage{7}
(\byear{2012})
\end{barticle}
\endbibitem

%%% 29
\bibitem[\protect\citeauthoryear{Tanaka et~al.}{2012}]{tanaka2012experimental}
\begin{barticle}
\bauthor{\bsnm{Tanaka}, \binits{Y.}},
\bauthor{\bsnm{Ren}, \binits{Z.}},
\bauthor{\bsnm{Sato}, \binits{T.}},
\bauthor{\bsnm{Nakayama}, \binits{K.}},
\bauthor{\bsnm{Souma}, \binits{S.}},
\bauthor{\bsnm{Takahashi}, \binits{T.}},
\bauthor{\bsnm{Segawa}, \binits{K.}},
\bauthor{\bsnm{Ando}, \binits{Y.}}:
\batitle{Experimental realization of a topological crystalline insulator in
  {S}n{T}e}.
\bjtitle{Nature Physics}
\bvolume{8}(\bissue{11}),
\bfpage{800}--\blpage{803}
(\byear{2012})
\end{barticle}
\endbibitem

%%% 30
\bibitem[\protect\citeauthoryear{Liu et~al.}{2013}]{liu2013two}
\begin{barticle}
\bauthor{\bsnm{Liu}, \binits{J.}},
\bauthor{\bsnm{Duan}, \binits{W.}},
\bauthor{\bsnm{Fu}, \binits{L.}}:
\batitle{Two types of surface states in topological crystalline insulators}.
\bjtitle{Physical Review B}
\bvolume{88}(\bissue{24}),
\bfpage{241303}
(\byear{2013})
\end{barticle}
\endbibitem

%%% 31
\bibitem[\protect\citeauthoryear{Wang et~al.}{2013}]{wang2013nontrivial}
\begin{barticle}
\bauthor{\bsnm{Wang}, \binits{Y.J.}},
\bauthor{\bsnm{Tsai}, \binits{W.-F.}},
\bauthor{\bsnm{Lin}, \binits{H.}},
\bauthor{\bsnm{Xu}, \binits{S.-Y.}},
\bauthor{\bsnm{Neupane}, \binits{M.}},
\bauthor{\bsnm{Hasan}, \binits{M.}},
\bauthor{\bsnm{Bansil}, \binits{A.}}:
\batitle{Nontrivial spin texture of the coaxial dirac cones on the surface of
  topological crystalline insulator {S}n{T}e}.
\bjtitle{Physical Review B}
\bvolume{87}(\bissue{23}),
\bfpage{235317}
(\byear{2013})
\end{barticle}
\endbibitem

%%% 32
\bibitem[\protect\citeauthoryear{Wang et~al.}{2016}]{wang2016electronic}
\begin{barticle}
\bauthor{\bsnm{Wang}, \binits{J.}},
\bauthor{\bsnm{Wang}, \binits{N.}},
\bauthor{\bsnm{Huang}, \binits{H.}},
\bauthor{\bsnm{Duan}, \binits{W.}}:
\batitle{Electronic properties of {S}n{T}e-class topological crystalline
  insulator materials}.
\bjtitle{Chinese Physics B}
\bvolume{25}(\bissue{11}),
\bfpage{117313}
(\byear{2016})
\end{barticle}
\endbibitem

%%% 33
\bibitem[\protect\citeauthoryear{Klett et~al.}{2018}]{klett2018proximity}
\begin{barticle}
\bauthor{\bsnm{Klett}, \binits{R.}},
\bauthor{\bsnm{Sch\"onle}, \binits{J.}},
\bauthor{\bsnm{Becker}, \binits{A.}},
\bauthor{\bsnm{Dyck}, \binits{D.}},
\bauthor{\bsnm{Borisov}, \binits{K.}},
\bauthor{\bsnm{Rott}, \binits{K.}},
\bauthor{\bsnm{Ramermann}, \binits{D.}},
\bauthor{\bsnm{B\"uker}, \binits{B.}},
\bauthor{\bsnm{Haskenhoff}, \binits{J.}},
\bauthor{\bsnm{Krieft}, \binits{J.}}, \betal:
\batitle{Proximity-induced superconductivity and quantum interference in
  topological crystalline insulator {S}n{T}e thin-film devices}.
\bjtitle{Nano Letters}
\bvolume{18}(\bissue{2}),
\bfpage{1264}--\blpage{1268}
(\byear{2018})
\end{barticle}
\endbibitem

%%% 34
\bibitem[\protect\citeauthoryear{Trimble et~al.}{2021}]{trimble2021josephson}
\begin{barticle}
\bauthor{\bsnm{Trimble}, \binits{C.}},
\bauthor{\bsnm{Wei}, \binits{M.}},
\bauthor{\bsnm{Yuan}, \binits{N.}},
\bauthor{\bsnm{Kalantre}, \binits{S.}},
\bauthor{\bsnm{Liu}, \binits{P.}},
\bauthor{\bsnm{Han}, \binits{H.-J.}},
\bauthor{\bsnm{Han}, \binits{M.-G.}},
\bauthor{\bsnm{Zhu}, \binits{Y.}},
\bauthor{\bsnm{Cha}, \binits{J.}},
\bauthor{\bsnm{Fu}, \binits{L.}}, \betal:
\batitle{Josephson detection of time-reversal symmetry broken superconductivity
  in {S}n{T}e nanowires}.
\bjtitle{npj Quantum Materials}
\bvolume{6}(\bissue{1}),
\bfpage{61}
(\byear{2021})
\end{barticle}
\endbibitem

%%% 35
\bibitem[\protect\citeauthoryear{Rachmilowitz
  et~al.}{2019}]{rachmilowitz2019proximity}
\begin{barticle}
\bauthor{\bsnm{Rachmilowitz}, \binits{B.}},
\bauthor{\bsnm{Zhao}, \binits{H.}},
\bauthor{\bsnm{Li}, \binits{H.}},
\bauthor{\bsnm{LaFleur}, \binits{A.}},
\bauthor{\bsnm{Schneeloch}, \binits{J.}},
\bauthor{\bsnm{Zhong}, \binits{R.}},
\bauthor{\bsnm{Gu}, \binits{G.}},
\bauthor{\bsnm{Zeljkovic}, \binits{I.}}:
\batitle{Proximity-induced superconductivity in a topological crystalline
  insulator}.
\bjtitle{Physical Review B}
\bvolume{100}(\bissue{24}),
\bfpage{241402}
(\byear{2019})
\end{barticle}
\endbibitem

%%% 36
\bibitem[\protect\citeauthoryear{Yang et~al.}{2019}]{yang2019superconductivity}
\begin{barticle}
\bauthor{\bsnm{Yang}, \binits{H.}},
\bauthor{\bsnm{Li}, \binits{Y.-Y.}},
\bauthor{\bsnm{Liu}, \binits{T.-T.}},
\bauthor{\bsnm{Xue}, \binits{H.-Y.}},
\bauthor{\bsnm{Guan}, \binits{D.-D.}},
\bauthor{\bsnm{Wang}, \binits{S.-Y.}},
\bauthor{\bsnm{Zheng}, \binits{H.}},
\bauthor{\bsnm{Liu}, \binits{C.-H.}},
\bauthor{\bsnm{Fu}, \binits{L.}},
\bauthor{\bsnm{Jia}, \binits{J.-F.}}:
\batitle{Superconductivity of topological surface states and strong proximity
  effect in {S}n$_{1-x}${P}b$_{x}${T}e--{P}b heterostructures}.
\bjtitle{Advanced Materials}
\bvolume{31}(\bissue{52}),
\bfpage{1905582}
(\byear{2019})
\end{barticle}
\endbibitem

%%% 37
\bibitem[\protect\citeauthoryear{Yang et~al.}{2020}]{yang2020multiple}
\begin{barticle}
\bauthor{\bsnm{Yang}, \binits{H.}},
\bauthor{\bsnm{Li}, \binits{Y.-Y.}},
\bauthor{\bsnm{Liu}, \binits{T.-T.}},
\bauthor{\bsnm{Guan}, \binits{D.-D.}},
\bauthor{\bsnm{Wang}, \binits{S.-Y.}},
\bauthor{\bsnm{Zheng}, \binits{H.}},
\bauthor{\bsnm{Liu}, \binits{C.}},
\bauthor{\bsnm{Fu}, \binits{L.}},
\bauthor{\bsnm{Jia}, \binits{J.-F.}}:
\batitle{Multiple in-gap states induced by topological surface states in the
  superconducting topological crystalline insulator heterostructure
  $\text{Sn}_{1-x}\text{Pb}_{x}\text{Te}-\text{Pb}$}.
\bjtitle{Physical Review Letters}
\bvolume{125}(\bissue{13}),
\bfpage{136802}
(\byear{2020})
\end{barticle}
\endbibitem

%%% 38
\bibitem[\protect\citeauthoryear{Liu et~al.}{2024}]{liu2024fermi}
\begin{barticle}
\bauthor{\bsnm{Liu}, \binits{T.}},
\bauthor{\bsnm{Yi}, \binits{Z.}},
\bauthor{\bsnm{Xie}, \binits{B.}},
\bauthor{\bsnm{Zheng}, \binits{W.}},
\bauthor{\bsnm{Guan}, \binits{D.}},
\bauthor{\bsnm{Wang}, \binits{S.}},
\bauthor{\bsnm{Zheng}, \binits{H.}},
\bauthor{\bsnm{Liu}, \binits{C.}},
\bauthor{\bsnm{Yang}, \binits{H.}},
\bauthor{\bsnm{Li}, \binits{Y.}}, \betal:
\batitle{Fermi level tuning in {S}n$_{1-x}${P}b$_x${T}e/{P}b heterostructure
  via changing interface roughness}.
\bjtitle{Science China Physics, Mechanics \& Astronomy}
\bvolume{67}(\bissue{8}),
\bfpage{286811}
(\byear{2024})
\end{barticle}
\endbibitem

%%% 39
\bibitem[\protect\citeauthoryear{Hulm et~al.}{1968}]{hulm1968superconducting}
\begin{barticle}
\bauthor{\bsnm{Hulm}, \binits{J.}},
\bauthor{\bsnm{Jones}, \binits{C.}},
\bauthor{\bsnm{Deis}, \binits{D.}},
\bauthor{\bsnm{Fairbank}, \binits{H.}},
\bauthor{\bsnm{Lawless}, \binits{P.}}:
\batitle{Superconducting interactions in {T}in {T}elluride}.
\bjtitle{Physical Review}
\bvolume{169}(\bissue{2}),
\bfpage{388}
(\byear{1968})
\end{barticle}
\endbibitem

%%% 40
\bibitem[\protect\citeauthoryear{Hein and Meijer}{1969}]{hein1969critical}
\begin{barticle}
\bauthor{\bsnm{Hein}, \binits{R.}},
\bauthor{\bsnm{Meijer}, \binits{P.}}:
\batitle{Critical magnetic fields of superconducting {S}n{T}e}.
\bjtitle{Physical Review}
\bvolume{179}(\bissue{2}),
\bfpage{497}
(\byear{1969})
\end{barticle}
\endbibitem

%%% 41
\bibitem[\protect\citeauthoryear{Erickson et~al.}{2009}]{erickson2009enhanced}
\begin{barticle}
\bauthor{\bsnm{Erickson}, \binits{A.}},
\bauthor{\bsnm{Chu}, \binits{J.-H.}},
\bauthor{\bsnm{Toney}, \binits{M.}},
\bauthor{\bsnm{Geballe}, \binits{T.}},
\bauthor{\bsnm{Fisher}, \binits{I.}}:
\batitle{Enhanced superconducting pairing interaction in indium-doped tin
  telluride}.
\bjtitle{Physical Review B}
\bvolume{79}(\bissue{2}),
\bfpage{024520}
(\byear{2009})
\end{barticle}
\endbibitem

%%% 42
\bibitem[\protect\citeauthoryear{Balakrishnan
  et~al.}{2013}]{balakrishnan2013superconducting}
\begin{barticle}
\bauthor{\bsnm{Balakrishnan}, \binits{G.}},
\bauthor{\bsnm{Bawden}, \binits{L.}},
\bauthor{\bsnm{Cavendish}, \binits{S.}},
\bauthor{\bsnm{Lees}, \binits{M.R.}}:
\batitle{Superconducting properties of the in-substituted topological
  crystalline insulator {S}n{T}e}.
\bjtitle{Physical Review B}
\bvolume{87}(\bissue{14}),
\bfpage{140507}
(\byear{2013})
\end{barticle}
\endbibitem

%%% 43
\bibitem[\protect\citeauthoryear{Zhong et~al.}{2013}]{zhong2013optimizing}
\begin{barticle}
\bauthor{\bsnm{Zhong}, \binits{R.}},
\bauthor{\bsnm{Schneeloch}, \binits{J.}},
\bauthor{\bsnm{Shi}, \binits{X.}},
\bauthor{\bsnm{Xu}, \binits{Z.}},
\bauthor{\bsnm{Zhang}, \binits{C.}},
\bauthor{\bsnm{Tranquada}, \binits{J.}},
\bauthor{\bsnm{Li}, \binits{Q.}},
\bauthor{\bsnm{Gu}, \binits{G.}}:
\batitle{Optimizing the superconducting transition temperature and upper
  critical field of {S}n$_{1-x}${I}n$_{x}${T}e}.
\bjtitle{Physical Review B}
\bvolume{88}(\bissue{2}),
\bfpage{020505}
(\byear{2013})
\end{barticle}
\endbibitem

%%% 44
\bibitem[\protect\citeauthoryear{Sato et~al.}{2013}]{sato2013fermiology}
\begin{barticle}
\bauthor{\bsnm{Sato}, \binits{T.}},
\bauthor{\bsnm{Tanaka}, \binits{Y.}},
\bauthor{\bsnm{Nakayama}, \binits{K.}},
\bauthor{\bsnm{Souma}, \binits{S.}},
\bauthor{\bsnm{Takahashi}, \binits{T.}},
\bauthor{\bsnm{Sasaki}, \binits{S.}},
\bauthor{\bsnm{Ren}, \binits{Z.}},
\bauthor{\bsnm{Taskin}, \binits{A.}},
\bauthor{\bsnm{Segawa}, \binits{K.}},
\bauthor{\bsnm{Ando}, \binits{Y.}}:
\batitle{Fermiology of the strongly spin-orbit coupled superconductor
  {S}n$_{1-x}${I}n$_{x}${T}e: Implications for topological superconductivity}.
\bjtitle{Physical Review Letters}
\bvolume{110}(\bissue{20}),
\bfpage{206804}
(\byear{2013})
\end{barticle}
\endbibitem

%%% 45
\bibitem[\protect\citeauthoryear{Maurya
  et~al.}{2014}]{maurya2014superconducting}
\begin{barticle}
\bauthor{\bsnm{Maurya}, \binits{V.}},
\bauthor{\bsnm{Srivastava}, \binits{P.}},
\bauthor{\bsnm{Patnaik}, \binits{S.}}, \betal:
\batitle{Superconducting properties of indium-doped topological crystalline
  insulator {S}n{T}e}.
\bjtitle{Europhysics Letters}
\bvolume{108}(\bissue{3}),
\bfpage{37010}
(\byear{2014})
\end{barticle}
\endbibitem

%%% 46
\bibitem[\protect\citeauthoryear{Maeda et~al.}{2017}]{maeda2017spin}
\begin{barticle}
\bauthor{\bsnm{Maeda}, \binits{S.}},
\bauthor{\bsnm{Hirose}, \binits{R.}},
\bauthor{\bsnm{Matano}, \binits{K.}},
\bauthor{\bsnm{Novak}, \binits{M.}},
\bauthor{\bsnm{Ando}, \binits{Y.}},
\bauthor{\bsnm{Zheng}, \binits{G.-q.}}:
\batitle{Spin-singlet superconductivity in the doped topological crystalline
  insulator {S}n$_{0.96}${I}n$_{0.04}${T}e}.
\bjtitle{Physical Review B}
\bvolume{96}(\bissue{10}),
\bfpage{104502}
(\byear{2017})
\end{barticle}
\endbibitem

%%% 47
\bibitem[\protect\citeauthoryear{Smylie
  et~al.}{2018}]{smylie2018superconductivity}
\begin{barticle}
\bauthor{\bsnm{Smylie}, \binits{M.}},
\bauthor{\bsnm{Claus}, \binits{H.}},
\bauthor{\bsnm{Kwok}, \binits{W.-K.}},
\bauthor{\bsnm{Louden}, \binits{E.}},
\bauthor{\bsnm{Eskildsen}, \binits{M.}},
\bauthor{\bsnm{Sefat}, \binits{A.}},
\bauthor{\bsnm{Zhong}, \binits{R.}},
\bauthor{\bsnm{Schneeloch}, \binits{J.}},
\bauthor{\bsnm{Gu}, \binits{G.}},
\bauthor{\bsnm{Bokari}, \binits{E.}}, \betal:
\batitle{Superconductivity, pairing symmetry, and disorder in the doped
  topological insulator {S}n$_{1-x}${I}n$_{x}${T}e te for x$\leq$ 0.10}.
\bjtitle{Physical Review B}
\bvolume{97}(\bissue{2}),
\bfpage{024511}
(\byear{2018})
\end{barticle}
\endbibitem

%%% 48
\bibitem[\protect\citeauthoryear{Bliesener
  et~al.}{2019}]{bliesener2019superconductivity}
\begin{barticle}
\bauthor{\bsnm{Bliesener}, \binits{A.}},
\bauthor{\bsnm{Feng}, \binits{J.}},
\bauthor{\bsnm{Taskin}, \binits{A.}},
\bauthor{\bsnm{Ando}, \binits{Y.}}:
\batitle{Superconductivity in {S}n$_{1-x}${I}n$_{x}${T}e thin films grown by
  molecular beam epitaxy}.
\bjtitle{Physical Review Materials}
\bvolume{3}(\bissue{10}),
\bfpage{101201}
(\byear{2019})
\end{barticle}
\endbibitem

%%% 49
\bibitem[\protect\citeauthoryear{Nomoto et~al.}{2020}]{nomoto2020microscopic}
\begin{barticle}
\bauthor{\bsnm{Nomoto}, \binits{T.}},
\bauthor{\bsnm{Kawamura}, \binits{M.}},
\bauthor{\bsnm{Koretsune}, \binits{T.}},
\bauthor{\bsnm{Arita}, \binits{R.}},
\bauthor{\bsnm{Machida}, \binits{T.}},
\bauthor{\bsnm{Hanaguri}, \binits{T.}},
\bauthor{\bsnm{Kriener}, \binits{M.}},
\bauthor{\bsnm{Taguchi}, \binits{Y.}},
\bauthor{\bsnm{Tokura}, \binits{Y.}}:
\batitle{Microscopic characterization of the superconducting gap function in
  {S}n$_{1-x}${I}n$_{x}${T}e}.
\bjtitle{Physical Review B}
\bvolume{101}(\bissue{1}),
\bfpage{014505}
(\byear{2020})
\end{barticle}
\endbibitem

%%% 50
\bibitem[\protect\citeauthoryear{Smylie et~al.}{2020}]{smylie2020nodeless}
\begin{barticle}
\bauthor{\bsnm{Smylie}, \binits{M.}},
\bauthor{\bsnm{Kobayashi}, \binits{K.}},
\bauthor{\bsnm{Takahashi}, \binits{T.}},
\bauthor{\bsnm{Chaparro}, \binits{C.}},
\bauthor{\bsnm{Snezhko}, \binits{A.}},
\bauthor{\bsnm{Kwok}, \binits{W.-K.}},
\bauthor{\bsnm{Welp}, \binits{U.}}:
\batitle{Nodeless superconducting gap in the candidate topological
  superconductor {S}n$_{1-x}${I}n$_{x}${T}e for $x= 0.7$}.
\bjtitle{Physical Review B}
\bvolume{101}(\bissue{9}),
\bfpage{094513}
(\byear{2020})
\end{barticle}
\endbibitem

%%% 51
\bibitem[\protect\citeauthoryear{Smylie et~al.}{2022}]{smylie2022full}
\begin{barticle}
\bauthor{\bsnm{Smylie}, \binits{M.}},
\bauthor{\bsnm{Kobayashi}, \binits{K.}},
\bauthor{\bsnm{Dans}, \binits{J.}},
\bauthor{\bsnm{Hebbeker}, \binits{H.}},
\bauthor{\bsnm{Chapai}, \binits{R.}},
\bauthor{\bsnm{Kwok}, \binits{W.-K.}},
\bauthor{\bsnm{Welp}, \binits{U.}}:
\batitle{Full superconducting gap in the candidate topological superconductor
  {I}n$_{1- x}${P}b$_x${T}e for $x= 0.2$}.
\bjtitle{Physical Review B}
\bvolume{106}(\bissue{5}),
\bfpage{054503}
(\byear{2022})
\end{barticle}
\endbibitem

%%% 52
\bibitem[\protect\citeauthoryear{Tewari and Sau}{2012}]{tewari2012topological}
\begin{barticle}
\bauthor{\bsnm{Tewari}, \binits{S.}},
\bauthor{\bsnm{Sau}, \binits{J.D.}}:
\batitle{Topological invariants for spin-orbit coupled superconductor
  nanowires}.
\bjtitle{Physical Review Letters}
\bvolume{109}(\bissue{15}),
\bfpage{150408}
(\byear{2012})
\end{barticle}
\endbibitem

%%% 53
\bibitem[\protect\citeauthoryear{Hell et~al.}{2017}]{hell2017two}
\begin{barticle}
\bauthor{\bsnm{Hell}, \binits{M.}},
\bauthor{\bsnm{Leijnse}, \binits{M.}},
\bauthor{\bsnm{Flensberg}, \binits{K.}}:
\batitle{Two-dimensional platform for networks of {M}ajorana bound states}.
\bjtitle{Physical Review Letters}
\bvolume{118}(\bissue{10}),
\bfpage{107701}
(\byear{2017})
\end{barticle}
\endbibitem

%%% 54
\bibitem[\protect\citeauthoryear{Sun and Jia}{2017}]{sun2017detection}
\begin{barticle}
\bauthor{\bsnm{Sun}, \binits{H.-H.}},
\bauthor{\bsnm{Jia}, \binits{J.-F.}}:
\batitle{Detection of {M}ajorana zero mode in the vortex}.
\bjtitle{npj Quantum Materials}
\bvolume{2}(\bissue{1}),
\bfpage{34}
(\byear{2017})
\end{barticle}
\endbibitem

%%% 55
\bibitem[\protect\citeauthoryear{Hosur et~al.}{2011}]{hosur2011majorana}
\begin{barticle}
\bauthor{\bsnm{Hosur}, \binits{P.}},
\bauthor{\bsnm{Ghaemi}, \binits{P.}},
\bauthor{\bsnm{Mong}, \binits{R.S.}},
\bauthor{\bsnm{Vishwanath}, \binits{A.}}:
\batitle{Majorana modes at the ends of superconductor vortices in doped
  topological insulators}.
\bjtitle{Physical Review Letters}
\bvolume{107}(\bissue{9}),
\bfpage{097001}
(\byear{2011})
\end{barticle}
\endbibitem

%%% 56
\bibitem[\protect\citeauthoryear{Chiu et~al.}{2011}]{chiu2011vortex}
\begin{barticle}
\bauthor{\bsnm{Chiu}, \binits{C.-K.}},
\bauthor{\bsnm{Gilbert}, \binits{M.J.}},
\bauthor{\bsnm{Hughes}, \binits{T.L.}}:
\batitle{Vortex lines in topological insulator-superconductor
  heterostructures}.
\bjtitle{Physical Review B}
\bvolume{84}(\bissue{14}),
\bfpage{144507}
(\byear{2011})
\end{barticle}
\endbibitem

%%% 57
\bibitem[\protect\citeauthoryear{Caroli et~al.}{1964}]{caroli1964bound}
\begin{barticle}
\bauthor{\bsnm{Caroli}, \binits{C.}},
\bauthor{\bsnm{De~Gennes}, \binits{P.}},
\bauthor{\bsnm{Matricon}, \binits{J.}}:
\batitle{Bound fermion states on a vortex line in a type ii superconductor}.
\bjtitle{Physics Letters}
\bvolume{9}(\bissue{4}),
\bfpage{307}--\blpage{309}
(\byear{1964})
\end{barticle}
\endbibitem

%%% 58
\bibitem[\protect\citeauthoryear{Volovik}{1999}]{volovik1999fermion}
\begin{barticle}
\bauthor{\bsnm{Volovik}, \binits{G.}}:
\batitle{Fermion zero modes on vortices in chiral superconductors}.
\bjtitle{Journal of Experimental and Theoretical Physics Letters}
\bvolume{70}(\bissue{9}),
\bfpage{609}--\blpage{614}
(\byear{1999})
\end{barticle}
\endbibitem

%%% 59
\bibitem[\protect\citeauthoryear{Khaymovich
  et~al.}{2009}]{khaymovich2009vortex}
\begin{barticle}
\bauthor{\bsnm{Khaymovich}, \binits{I.}},
\bauthor{\bsnm{Kopnin}, \binits{N.}},
\bauthor{\bsnm{Mel’Nikov}, \binits{A.}},
\bauthor{\bsnm{Shereshevskii}, \binits{I.}}:
\batitle{Vortex core states in superconducting graphene}.
\bjtitle{Physical Review B}
\bvolume{79}(\bissue{22}),
\bfpage{224506}
(\byear{2009})
\end{barticle}
\endbibitem

%%% 60
\bibitem[\protect\citeauthoryear{Hess et~al.}{1990}]{hess1990vortex}
\begin{barticle}
\bauthor{\bsnm{Hess}, \binits{H.}},
\bauthor{\bsnm{Robinson}, \binits{R.}},
\bauthor{\bsnm{Waszczak}, \binits{J.}}:
\batitle{Vortex-core structure observed with a scanning tunneling microscope}.
\bjtitle{Physical Review Letters}
\bvolume{64}(\bissue{22}),
\bfpage{2711}
(\byear{1990})
\end{barticle}
\endbibitem

%%% 61
\bibitem[\protect\citeauthoryear{Li et~al.}{2014}]{li2014majorana}
\begin{barticle}
\bauthor{\bsnm{Li}, \binits{Z.-Z.}},
\bauthor{\bsnm{Zhang}, \binits{F.-C.}},
\bauthor{\bsnm{Wang}, \binits{Q.-H.}}:
\batitle{Majorana modes in a topological insulator/s-wave superconductor
  heterostructure}.
\bjtitle{Scientific Reports}
\bvolume{4}(\bissue{1}),
\bfpage{1}--\blpage{6}
(\byear{2014})
\end{barticle}
\endbibitem

%%% 62
\bibitem[\protect\citeauthoryear{Kawakami and Hu}{2015}]{kawakami2015evolution}
\begin{barticle}
\bauthor{\bsnm{Kawakami}, \binits{T.}},
\bauthor{\bsnm{Hu}, \binits{X.}}:
\batitle{Evolution of density of states and a spin-resolved checkerboard-type
  pattern associated with the {M}ajorana bound state}.
\bjtitle{Physical Review Letters}
\bvolume{115}(\bissue{17}),
\bfpage{177001}
(\byear{2015})
\end{barticle}
\endbibitem

%%% 63
\bibitem[\protect\citeauthoryear{Yan et~al.}{2020}]{yan2020vortex}
\begin{barticle}
\bauthor{\bsnm{Yan}, \binits{Z.}},
\bauthor{\bsnm{Wu}, \binits{Z.}},
\bauthor{\bsnm{Huang}, \binits{W.}}:
\batitle{Vortex end {M}ajorana zero modes in superconducting {D}irac and {W}eyl
  semimetals}.
\bjtitle{Physical Review Letters}
\bvolume{124}(\bissue{25}),
\bfpage{257001}
(\byear{2020})
\end{barticle}
\endbibitem

%%% 64
\bibitem[\protect\citeauthoryear{Kobayashi and
  Furusaki}{2020}]{kobayashi2020double}
\begin{barticle}
\bauthor{\bsnm{Kobayashi}, \binits{S.}},
\bauthor{\bsnm{Furusaki}, \binits{A.}}:
\batitle{Double {M}ajorana vortex zero modes in superconducting topological
  crystalline insulators with surface rotation anomaly}.
\bjtitle{Physical Review B}
\bvolume{102}(\bissue{18}),
\bfpage{180505}
(\byear{2020})
\end{barticle}
\endbibitem

%%% 65
\bibitem[\protect\citeauthoryear{Kobayashi et~al.}{2023}]{kobayashi2023crystal}
\begin{barticle}
\bauthor{\bsnm{Kobayashi}, \binits{S.}},
\bauthor{\bsnm{Sumita}, \binits{S.}},
\bauthor{\bsnm{Hirayama}, \binits{M.}},
\bauthor{\bsnm{Furusaki}, \binits{A.}}:
\batitle{Crystal-symmetry-protected gapless vortex-line phases in
  superconducting {D}irac semimetals}.
\bjtitle{Physical Review B}
\bvolume{107}(\bissue{21}),
\bfpage{214518}
(\byear{2023})
\end{barticle}
\endbibitem

%%% 66
\bibitem[\protect\citeauthoryear{Hu et~al.}{2022}]{hu2022competing}
\begin{barticle}
\bauthor{\bsnm{Hu}, \binits{L.-H.}},
\bauthor{\bsnm{Wu}, \binits{X.}},
\bauthor{\bsnm{Liu}, \binits{C.-X.}},
\bauthor{\bsnm{Zhang}, \binits{R.-X.}}:
\batitle{Competing vortex topologies in iron-based superconductors}.
\bjtitle{Physical Review Letters}
\bvolume{129}(\bissue{27}),
\bfpage{277001}
(\byear{2022})
\end{barticle}
\endbibitem

%%% 67
\bibitem[\protect\citeauthoryear{Altland and
  Zirnbauer}{1997}]{altland1997nonstandard}
\begin{barticle}
\bauthor{\bsnm{Altland}, \binits{A.}},
\bauthor{\bsnm{Zirnbauer}, \binits{M.R.}}:
\batitle{Nonstandard symmetry classes in mesoscopic normal-superconducting
  hybrid structures}.
\bjtitle{Physical Review B}
\bvolume{55}(\bissue{2}),
\bfpage{1142}
(\byear{1997})
\end{barticle}
\endbibitem

%%% 68
\bibitem[\protect\citeauthoryear{Schnyder
  et~al.}{2008}]{schnyder2008classification}
\begin{barticle}
\bauthor{\bsnm{Schnyder}, \binits{A.P.}},
\bauthor{\bsnm{Ryu}, \binits{S.}},
\bauthor{\bsnm{Furusaki}, \binits{A.}},
\bauthor{\bsnm{Ludwig}, \binits{A.W.}}:
\batitle{Classification of topological insulators and superconductors in three
  spatial dimensions}.
\bjtitle{Physical Review B}
\bvolume{78}(\bissue{19}),
\bfpage{195125}
(\byear{2008})
\end{barticle}
\endbibitem

%%% 69
\bibitem[\protect\citeauthoryear{Chiu et~al.}{2016}]{chiu2016classification}
\begin{barticle}
\bauthor{\bsnm{Chiu}, \binits{C.-K.}},
\bauthor{\bsnm{Teo}, \binits{J.C.}},
\bauthor{\bsnm{Schnyder}, \binits{A.P.}},
\bauthor{\bsnm{Ryu}, \binits{S.}}:
\batitle{Classification of topological quantum matter with symmetries}.
\bjtitle{Reviews of Modern Physics}
\bvolume{88}(\bissue{3}),
\bfpage{035005}
(\byear{2016})
\end{barticle}
\endbibitem

%%% 70
\bibitem[\protect\citeauthoryear{Wei{\ss}e et~al.}{2006}]{weisse2006kernel}
\begin{barticle}
\bauthor{\bsnm{Wei{\ss}e}, \binits{A.}},
\bauthor{\bsnm{Wellein}, \binits{G.}},
\bauthor{\bsnm{Alvermann}, \binits{A.}},
\bauthor{\bsnm{Fehske}, \binits{H.}}:
\batitle{The kernel polynomial method}.
\bjtitle{Reviews of modern physics}
\bvolume{78}(\bissue{1}),
\bfpage{275}
(\byear{2006})
\end{barticle}
\endbibitem

%%% 71
\bibitem[\protect\citeauthoryear{Nagai et~al.}{2012}]{nagai2012efficient}
\begin{barticle}
\bauthor{\bsnm{Nagai}, \binits{Y.}},
\bauthor{\bsnm{Ota}, \binits{Y.}},
\bauthor{\bsnm{Machida}, \binits{M.}}:
\batitle{Efficient numerical self-consistent mean-field approach for fermionic
  many-body systems by polynomial expansion on spectral density}.
\bjtitle{Journal of the Physical Society of Japan}
\bvolume{81}(\bissue{2}),
\bfpage{024710}
(\byear{2012})
\end{barticle}
\endbibitem

%%% 72
\bibitem[\protect\citeauthoryear{Xiong et~al.}{2017}]{xiong2017anisotropic}
\begin{barticle}
\bauthor{\bsnm{Xiong}, \binits{Y.}},
\bauthor{\bsnm{Yamakage}, \binits{A.}},
\bauthor{\bsnm{Kobayashi}, \binits{S.}},
\bauthor{\bsnm{Sato}, \binits{M.}},
\bauthor{\bsnm{Tanaka}, \binits{Y.}}:
\batitle{Anisotropic magnetic responses of topological crystalline
  superconductors}.
\bjtitle{Crystals}
\bvolume{7}(\bissue{2}),
\bfpage{58}
(\byear{2017})
\end{barticle}
\endbibitem

%%% 73
\bibitem[\protect\citeauthoryear{Kobayashi
  et~al.}{2019}]{kobayashi2019majorana}
\begin{barticle}
\bauthor{\bsnm{Kobayashi}, \binits{S.}},
\bauthor{\bsnm{Yamakage}, \binits{A.}},
\bauthor{\bsnm{Tanaka}, \binits{Y.}},
\bauthor{\bsnm{Sato}, \binits{M.}}:
\batitle{Majorana multipole response of topological superconductors}.
\bjtitle{Physical Review Letters}
\bvolume{123}(\bissue{9}),
\bfpage{097002}
(\byear{2019})
\end{barticle}
\endbibitem

%%% 74
\bibitem[\protect\citeauthoryear{Yamazaki et~al.}{2021}]{yamazaki2021magnetic}
\begin{barticle}
\bauthor{\bsnm{Yamazaki}, \binits{Y.}},
\bauthor{\bsnm{Kobayashi}, \binits{S.}},
\bauthor{\bsnm{Yamakage}, \binits{A.}}:
\batitle{Magnetic response of {M}ajorana {K}ramers pairs with an order-two
  symmetry}.
\bjtitle{Physical Review B}
\bvolume{103}(\bissue{9}),
\bfpage{094508}
(\byear{2021})
\end{barticle}
\endbibitem

%%% 75
\bibitem[\protect\citeauthoryear{Kobayashi
  et~al.}{2021}]{kobayashi2021majorana}
\begin{barticle}
\bauthor{\bsnm{Kobayashi}, \binits{S.}},
\bauthor{\bsnm{Yamazaki}, \binits{Y.}},
\bauthor{\bsnm{Yamakage}, \binits{A.}},
\bauthor{\bsnm{Sato}, \binits{M.}}:
\batitle{Majorana multipole response: {G}eneral theory and application to
  wallpaper groups}.
\bjtitle{Physical Review B}
\bvolume{103}(\bissue{22}),
\bfpage{224504}
(\byear{2021})
\end{barticle}
\endbibitem

%%% 76
\bibitem[\protect\citeauthoryear{Kobayashi and
  Sato}{2024}]{kobayashi2024electromagnetic}
\begin{barticle}
\bauthor{\bsnm{Kobayashi}, \binits{S.}},
\bauthor{\bsnm{Sato}, \binits{M.}}:
\batitle{Electromagnetic response of spinful {M}ajorana fermions}.
\bjtitle{Progress of Theoretical and Experimental Physics}
\bvolume{2024}(\bissue{8}),
\bfpage{08}--\blpage{106}
(\byear{2024})
\end{barticle}
\endbibitem

%%% 77
\bibitem[\protect\citeauthoryear{Yamazaki and
  Kobayashi}{2024}]{yamazaki2024majorana}
\begin{botherref}
\oauthor{\bsnm{Yamazaki}, \binits{Y.}},
\oauthor{\bsnm{Kobayashi}, \binits{S.}}:
Majorana multipole response with magnetic point group symmetry.
arXiv preprint arXiv:2407.01924
(2024)
\end{botherref}
\endbibitem

%%% 78
\bibitem[\protect\citeauthoryear{Liu et~al.}{2024}]{liu2024signatures}
\begin{barticle}
\bauthor{\bsnm{Liu}, \binits{T.}},
\bauthor{\bsnm{Wan}, \binits{C.Y.}},
\bauthor{\bsnm{Yang}, \binits{H.}},
\bauthor{\bsnm{Zhao}, \binits{Y.}},
\bauthor{\bsnm{Xie}, \binits{B.}},
\bauthor{\bsnm{Zheng}, \binits{W.}},
\bauthor{\bsnm{Yi}, \binits{Z.}},
\bauthor{\bsnm{Guan}, \binits{D.}},
\bauthor{\bsnm{Wang}, \binits{S.}},
\bauthor{\bsnm{Zheng}, \binits{H.}}, \betal:
\batitle{Signatures of hybridization of multiple {M}ajorana zero modes in a
  vortex}.
\bjtitle{Nature}
\bvolume{633},
\bfpage{71}
(\byear{2024})
\end{barticle}
\endbibitem

%%% 79
\bibitem[\protect\citeauthoryear{Shore et~al.}{1989}]{shore1989density}
\begin{barticle}
\bauthor{\bsnm{Shore}, \binits{J.D.}},
\bauthor{\bsnm{Huang}, \binits{M.}},
\bauthor{\bsnm{Dorsey}, \binits{A.T.}},
\bauthor{\bsnm{Sethna}, \binits{J.P.}}:
\batitle{Density of states in a vortex core and the zero-bias tunneling peak}.
\bjtitle{Physical review letters}
\bvolume{62}(\bissue{26}),
\bfpage{3089}
(\byear{1989})
\end{barticle}
\endbibitem

%%% 80
\bibitem[\protect\citeauthoryear{Gygi and Schl{\"u}ter}{1991}]{gygi1991self}
\begin{barticle}
\bauthor{\bsnm{Gygi}, \binits{F.}},
\bauthor{\bsnm{Schl{\"u}ter}, \binits{M.}}:
\batitle{Self-consistent electronic structure of a vortex line in a type-{I}{I}
  superconductor}.
\bjtitle{Physical Review B}
\bvolume{43}(\bissue{10}),
\bfpage{7609}
(\byear{1991})
\end{barticle}
\endbibitem

%%% 81
\bibitem[\protect\citeauthoryear{Datta}{1997}]{datta1997electronic}
\begin{bbook}
\bauthor{\bsnm{Datta}, \binits{S.}}:
\bbtitle{Electronic Transport in Mesoscopic Systems}.
\bpublisher{Cambridge university press},
\blocation{Cambridge}
(\byear{1997})
\end{bbook}
\endbibitem

%%% 82
\bibitem[\protect\citeauthoryear{Nagai et~al.}{2012}]{nagai2012direct}
\begin{barticle}
\bauthor{\bsnm{Nagai}, \binits{Y.}},
\bauthor{\bsnm{Nakai}, \binits{N.}},
\bauthor{\bsnm{Machida}, \binits{M.}}:
\batitle{Direct numerical demonstration of sign-preserving quasiparticle
  interference via an impurity inside a vortex core in an unconventional
  superconductor}.
\bjtitle{Physical Review B}
\bvolume{85}(\bissue{9}),
\bfpage{092505}
(\byear{2012})
\end{barticle}
\endbibitem

%%% 83
\bibitem[\protect\citeauthoryear{Smith et~al.}{2016}]{smith2016manifestation}
\begin{barticle}
\bauthor{\bsnm{Smith}, \binits{E.D.}},
\bauthor{\bsnm{Tanaka}, \binits{K.}},
\bauthor{\bsnm{Nagai}, \binits{Y.}}:
\batitle{Manifestation of chirality in the vortex lattice in a two-dimensional
  topological superconductor}.
\bjtitle{Physical Review B}
\bvolume{94}(\bissue{6}),
\bfpage{064515}
(\byear{2016})
\end{barticle}
\endbibitem

%%% 84
\bibitem[\protect\citeauthoryear{Berthod}{2016}]{berthod2016vortex}
\begin{barticle}
\bauthor{\bsnm{Berthod}, \binits{C.}}:
\batitle{Vortex spectroscopy in the vortex glass: {A} real-space numerical
  approach}.
\bjtitle{Physical Review B}
\bvolume{94}(\bissue{18}),
\bfpage{184510}
(\byear{2016})
\end{barticle}
\endbibitem

%%% 85
\bibitem[\protect\citeauthoryear{Galvis et~al.}{2018}]{galvis2018tilted}
\begin{barticle}
\bauthor{\bsnm{Galvis}, \binits{J.}},
\bauthor{\bsnm{Herrera}, \binits{E.}},
\bauthor{\bsnm{Berthod}, \binits{C.}},
\bauthor{\bsnm{Vieira}, \binits{S.}},
\bauthor{\bsnm{Guillam{\'o}n}, \binits{I.}},
\bauthor{\bsnm{Suderow}, \binits{H.}}:
\batitle{Tilted vortex cores and superconducting gap anisotropy in
  2{H}-{N}b{S}e$_2$}.
\bjtitle{Communications Physics}
\bvolume{1}(\bissue{1}),
\bfpage{1}--\blpage{9}
(\byear{2018})
\end{barticle}
\endbibitem

%%% 86
\bibitem[\protect\citeauthoryear{Berthod}{2018}]{berthod2018signatures}
\begin{barticle}
\bauthor{\bsnm{Berthod}, \binits{C.}}:
\batitle{Signatures of nodeless multiband superconductivity and particle-hole
  crossover in the vortex cores of {F}e{T}e$_{0.55}${S}e$_{0.45}$}.
\bjtitle{Physical Review B}
\bvolume{98}(\bissue{14}),
\bfpage{144519}
(\byear{2018})
\end{barticle}
\endbibitem

%%% 87
\bibitem[\protect\citeauthoryear{Mitchell and
  Wallis}{1966}]{mitchell1966theoretical}
\begin{barticle}
\bauthor{\bsnm{Mitchell}, \binits{D.}},
\bauthor{\bsnm{Wallis}, \binits{R.}}:
\batitle{Theoretical energy-band parameters for the lead salts}.
\bjtitle{Physical Review}
\bvolume{151}(\bissue{2}),
\bfpage{581}
(\byear{1966})
\end{barticle}
\endbibitem

%%% 88
\bibitem[\protect\citeauthoryear{Fulga et~al.}{2016}]{fulga2016coupled}
\begin{barticle}
\bauthor{\bsnm{Fulga}, \binits{I.}},
\bauthor{\bsnm{Avraham}, \binits{N.}},
\bauthor{\bsnm{Beidenkopf}, \binits{H.}},
\bauthor{\bsnm{Stern}, \binits{A.}}:
\batitle{Coupled-layer description of topological crystalline insulators}.
\bjtitle{Physical Review B}
\bvolume{94}(\bissue{12}),
\bfpage{125405}
(\byear{2016})
\end{barticle}
\endbibitem

%%% 89
\bibitem[\protect\citeauthoryear{Zhu et~al.}{2021}]{zhu2021discovery}
\begin{barticle}
\bauthor{\bsnm{Zhu}, \binits{Z.}},
\bauthor{\bsnm{Papaj}, \binits{M.}},
\bauthor{\bsnm{Nie}, \binits{X.-A.}},
\bauthor{\bsnm{Xu}, \binits{H.-K.}},
\bauthor{\bsnm{Gu}, \binits{Y.-S.}},
\bauthor{\bsnm{Yang}, \binits{X.}},
\bauthor{\bsnm{Guan}, \binits{D.}},
\bauthor{\bsnm{Wang}, \binits{S.}},
\bauthor{\bsnm{Li}, \binits{Y.}},
\bauthor{\bsnm{Liu}, \binits{C.}}, \betal:
\batitle{Discovery of segmented fermi surface induced by {C}ooper pair
  momentum}.
\bjtitle{Science}
\bvolume{374}(\bissue{6573}),
\bfpage{1381}--\blpage{1385}
(\byear{2021})
\end{barticle}
\endbibitem

%%% 90
\bibitem[\protect\citeauthoryear{Pan et~al.}{2024}]{pan2024majorana}
\begin{barticle}
\bauthor{\bsnm{Pan}, \binits{X.-H.}},
\bauthor{\bsnm{Chen}, \binits{L.}},
\bauthor{\bsnm{Liu}, \binits{D.E.}},
\bauthor{\bsnm{Zhang}, \binits{F.-C.}},
\bauthor{\bsnm{Liu}, \binits{X.}}:
\batitle{Majorana zero modes induced by the {M}eissner effect at small magnetic
  field}.
\bjtitle{Physical Review Letters}
\bvolume{132}(\bissue{3}),
\bfpage{036602}
(\byear{2024})
\end{barticle}
\endbibitem

%%% 91
\bibitem[\protect\citeauthoryear{Po et~al.}{2017}]{po2017symmetry}
\begin{barticle}
\bauthor{\bsnm{Po}, \binits{H.C.}},
\bauthor{\bsnm{Vishwanath}, \binits{A.}},
\bauthor{\bsnm{Watanabe}, \binits{H.}}:
\batitle{Symmetry-based indicators of band topology in the 230 space groups}.
\bjtitle{Nature communications}
\bvolume{8}(\bissue{1}),
\bfpage{50}
(\byear{2017})
\end{barticle}
\endbibitem

%%% 92
\bibitem[\protect\citeauthoryear{Bradlyn et~al.}{2017}]{bradlyn2017topological}
\begin{barticle}
\bauthor{\bsnm{Bradlyn}, \binits{B.}},
\bauthor{\bsnm{Elcoro}, \binits{L.}},
\bauthor{\bsnm{Cano}, \binits{J.}},
\bauthor{\bsnm{Vergniory}, \binits{M.G.}},
\bauthor{\bsnm{Wang}, \binits{Z.}},
\bauthor{\bsnm{Felser}, \binits{C.}},
\bauthor{\bsnm{Aroyo}, \binits{M.I.}},
\bauthor{\bsnm{Bernevig}, \binits{B.A.}}:
\batitle{Topological quantum chemistry}.
\bjtitle{Nature}
\bvolume{547}(\bissue{7663}),
\bfpage{298}--\blpage{305}
(\byear{2017})
\end{barticle}
\endbibitem

%%% 93
\bibitem[\protect\citeauthoryear{Kruthoff
  et~al.}{2017}]{kruthoff2017topological}
\begin{barticle}
\bauthor{\bsnm{Kruthoff}, \binits{J.}},
\bauthor{\bsnm{De~Boer}, \binits{J.}},
\bauthor{\bsnm{Van~Wezel}, \binits{J.}},
\bauthor{\bsnm{Kane}, \binits{C.L.}},
\bauthor{\bsnm{Slager}, \binits{R.-J.}}:
\batitle{Topological classification of crystalline insulators through band
  structure combinatorics}.
\bjtitle{Physical Review X}
\bvolume{7}(\bissue{4}),
\bfpage{041069}
(\byear{2017})
\end{barticle}
\endbibitem

%%% 94
\bibitem[\protect\citeauthoryear{Ono et~al.}{2019}]{ono2019symmetry}
\begin{barticle}
\bauthor{\bsnm{Ono}, \binits{S.}},
\bauthor{\bsnm{Yanase}, \binits{Y.}},
\bauthor{\bsnm{Watanabe}, \binits{H.}}:
\batitle{Symmetry indicators for topological superconductors}.
\bjtitle{Physical Review Research}
\bvolume{1}(\bissue{1}),
\bfpage{013012}
(\byear{2019})
\end{barticle}
\endbibitem

%%% 95
\bibitem[\protect\citeauthoryear{Ono et~al.}{2021}]{ono2021z}
\begin{barticle}
\bauthor{\bsnm{Ono}, \binits{S.}},
\bauthor{\bsnm{Po}, \binits{H.C.}},
\bauthor{\bsnm{Shiozaki}, \binits{K.}}:
\batitle{Z 2-enriched symmetry indicators for topological superconductors in
  the 1651 magnetic space groups}.
\bjtitle{Physical Review Research}
\bvolume{3}(\bissue{2}),
\bfpage{023086}
(\byear{2021})
\end{barticle}
\endbibitem

%%% 96
\bibitem[\protect\citeauthoryear{Qin et~al.}{2019}]{qin2019quasi}
\begin{barticle}
\bauthor{\bsnm{Qin}, \binits{S.}},
\bauthor{\bsnm{Hu}, \binits{L.}},
\bauthor{\bsnm{Le}, \binits{C.}},
\bauthor{\bsnm{Zeng}, \binits{J.}},
\bauthor{\bsnm{Zhang}, \binits{F.-c.}},
\bauthor{\bsnm{Fang}, \binits{C.}},
\bauthor{\bsnm{Hu}, \binits{J.}}:
\batitle{Quasi-1d topological nodal vortex line phase in doped superconducting
  3d dirac semimetals}.
\bjtitle{Physical Review Letters}
\bvolume{123}(\bissue{2}),
\bfpage{027003}
(\byear{2019})
\end{barticle}
\endbibitem

%%% 97
\bibitem[\protect\citeauthoryear{Jackiw and Rossi}{1981}]{jackiw1981zero}
\begin{barticle}
\bauthor{\bsnm{Jackiw}, \binits{R.}},
\bauthor{\bsnm{Rossi}, \binits{P.}}:
\batitle{Zero modes of the vortex-fermion system}.
\bjtitle{Nuclear Physics B}
\bvolume{190}(\bissue{4}),
\bfpage{681}--\blpage{691}
(\byear{1981})
\end{barticle}
\endbibitem

%%% 98
\bibitem[\protect\citeauthoryear{Sancho et~al.}{1985}]{sancho1985highly}
\begin{barticle}
\bauthor{\bsnm{Sancho}, \binits{M.L.}},
\bauthor{\bsnm{Sancho}, \binits{J.L.}},
\bauthor{\bsnm{Sancho}, \binits{J.L.}},
\bauthor{\bsnm{Rubio}, \binits{J.}}:
\batitle{Highly convergent schemes for the calculation of bulk and surface
  {G}reen functions}.
\bjtitle{Journal of Physics F: Metal Physics}
\bvolume{15}(\bissue{4}),
\bfpage{851}
(\byear{1985})
\end{barticle}
\endbibitem

%%% 99
\bibitem[\protect\citeauthoryear{Pahomi et~al.}{2020}]{pahomi2020braiding}
\begin{barticle}
\bauthor{\bsnm{Pahomi}, \binits{T.E.}},
\bauthor{\bsnm{Sigrist}, \binits{M.}},
\bauthor{\bsnm{Soluyanov}, \binits{A.A.}}:
\batitle{Braiding {M}ajorana corner modes in a second-order topological
  superconductor}.
\bjtitle{Physical Review Research}
\bvolume{2}(\bissue{3}),
\bfpage{032068}
(\byear{2020})
\end{barticle}
\endbibitem

%%% 100
\bibitem[\protect\citeauthoryear{Pan et~al.}{2022}]{pan2022detecting}
\begin{barticle}
\bauthor{\bsnm{Pan}, \binits{X.-H.}},
\bauthor{\bsnm{Luo}, \binits{X.-J.}},
\bauthor{\bsnm{Gao}, \binits{J.-H.}},
\bauthor{\bsnm{Liu}, \binits{X.}}:
\batitle{Detecting and braiding higher-order {M}ajorana corner states through
  their spin degree of freedom}.
\bjtitle{Physical Review B}
\bvolume{105}(\bissue{19}),
\bfpage{195106}
(\byear{2022})
\end{barticle}
\endbibitem

%%% 101
\bibitem[\protect\citeauthoryear{Liu et~al.}{2024}]{liu2024magnetically}
\begin{barticle}
\bauthor{\bsnm{Liu}, \binits{L.}},
\bauthor{\bsnm{Miao}, \binits{C.}},
\bauthor{\bsnm{Tang}, \binits{H.}},
\bauthor{\bsnm{Zhang}, \binits{Y.-T.}},
\bauthor{\bsnm{Qiao}, \binits{Z.}}:
\batitle{Magnetically controlled topological braiding with majorana corner
  states in second-order topological superconductors}.
\bjtitle{Physical Review B}
\bvolume{109}(\bissue{11}),
\bfpage{115413}
(\byear{2024})
\end{barticle}
\endbibitem

%%% 102
\bibitem[\protect\citeauthoryear{Sato and Fujimoto}{2016}]{sato2016majorana}
\begin{barticle}
\bauthor{\bsnm{Sato}, \binits{M.}},
\bauthor{\bsnm{Fujimoto}, \binits{S.}}:
\batitle{Majorana fermions and topology in superconductors}.
\bjtitle{Journal of the Physical Society of Japan}
\bvolume{85}(\bissue{7}),
\bfpage{072001}
(\byear{2016})
\end{barticle}
\endbibitem

\end{thebibliography}
\end{document}